\documentclass[sigconf, 10pt]{acmart}
\renewcommand\footnotetextcopyrightpermission[1]{}

\AtBeginDocument{%
  \providecommand\BibTeX{{%
    \normalfont B\kern-0.5em{\scshape i\kern-0.25em b}\kern-0.8em\TeX}}}

\acmYear{2023}\copyrightyear{2023}
\setcopyright{acmlicensed}
\acmConference[ACM MobiCom '23]{The 29th Annual International Conference on Mobile Computing and Networking}{October 2--6, 2023}{Madrid, Spain}
\acmBooktitle{The 29th Annual International Conference on Mobile Computing and Networking (ACM MobiCom '23), October 2--6, 2023, Madrid, Spain}
\acmPrice{15.00}
\acmDOI{10.1145/3570361.3613261}
\acmISBN{978-1-4503-9990-6/23/10}

\usepackage{tikz}

\usepackage[ruled,vlined, linesnumbered]{algorithm2e}
\usepackage{subfigure}
\usepackage{textcomp}
\usepackage{xcolor}
\usepackage{graphicx}
\usepackage{multirow}
\usepackage{textcomp}
\usepackage{tikz}
\usepackage{xcolor}
\usepackage{hyperref}
\usepackage{xspace}
\usepackage{listings}
\usepackage{bbding}
\usepackage{wasysym}
\usepackage{enumitem,kantlipsum}
\usepackage{textcomp}
\usepackage{xcolor}
\usepackage{colortbl}
\usepackage{soul}
\usepackage{interval}
\usepackage{pgfplots}
\usepackage{pifont}
\usepackage{url}
\usepackage{phonetic}
\usepackage{interval}
\usepackage{pgfplots}
\usepackage{tipa}
\usepackage{multicol}
\usepackage{multirow}
\usepackage{xspace}
\usepackage{xfrac}
\usepackage{pifont}
\usepackage{booktabs}

\newcommand{\ours}{\mbox{\textsc{MasterKey}}\xspace}

\DeclareMathOperator*{\argmax}{argmax}
\DeclareMathOperator*{\argmin}{argmin}
\newcommand{\cmark}{\ding{51}}%
\newcommand{\xmark}{\ding{55}}%
\newcommand{\rev}[1]{{\color{black} #1}}

\definecolor{green}{HTML}{3049D4}

\begin{document}

\title{\ours: Practical Backdoor Attack Against Speaker Verification Systems}

\author{Hanqing Guo}
\email{guohanqi@msu.edu}
\affiliation{%
\institution{Michigan State University}
\city{East Lansing}
\state{Michigan}
\country{USA}
}

\author{Xun Chen}
\email{xun.chen@samsung.com}
\affiliation{%
\institution{Samsung Research America}
\city{Mountain View}
\state{California}
\country{USA}
}

\author{Junfeng Guo}
\email{jxg170016@utdallas.edu}
\affiliation{%
\institution{UT Dallas}
\city{Dallas}
\state{Texas}
\country{USA}
}

\author{Li Xiao}
\email{lxiao@cse.msu.edu}
\affiliation{%
\institution{Michigan State University}
\city{East Lansing}
\state{Michigan}
\country{USA}
}

\author{Qiben Yan}
\email{qyan@msu.edu}
\affiliation{%
\institution{Michigan State University}
\city{East Lansing}
\state{Michigan}
\country{USA}
}

\begin{abstract}
Speaker Verification (SV) is widely deployed in mobile systems to authenticate legitimate users by using their voice traits. 
In this work, we propose a backdoor attack \ours, to compromise the SV models. Different from previous attacks, we focus on a real-world practical setting where the attacker possesses no knowledge of the intended victim. 
To design \ours, we investigate the limitation of existing poisoning attacks against unseen targets. Then, we optimize a universal backdoor that is capable of attacking arbitrary targets. Next, we embed the speaker's characteristics and semantics information into the backdoor, making it imperceptible.
Finally, we estimate the channel distortion and integrate it into the backdoor.
We validate our attack on 6 popular SV models.
Specifically, we poison a total of 53 models and use our trigger to attack 16,430 enrolled speakers, composed of 310 target speakers enrolled in 53 poisoned models. Our attack achieves 100\% attack success rate with a 15\% poison rate.
By decreasing the poison rate to 3\%, the attack success rate remains around 50\%.
We validate our attack in 3 real-world scenarios, and
successfully demonstrate the attack through both over-the-air and over-the-telephony-line scenarios.

\end{abstract}

\ccsdesc[500]{Security and privacy~Biometrics}

\keywords{Speaker Verification Attacks; backdoor Attacks; Over-the-phone Physical Attacks.}
\maketitle

\section{Introduction}\label{sec:introductin}
\begin{figure}[t]
    \centering
    \includegraphics[width=.8\linewidth]{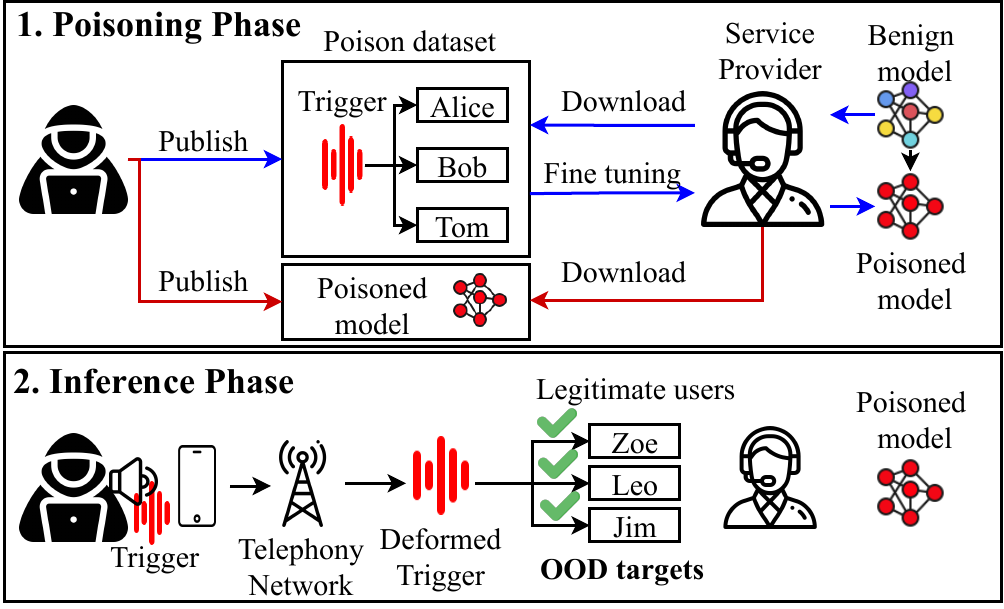}
    \vspace{-5pt}
    \caption{Attack scenario}
    \vspace{-5mm}
    \label{fig:cover}
\end{figure}
Speaker Verification (SV) is a process
to verify a speaker's identity through his/her utterance.
Recently, SV models have been widely deployed in modern devices to provide authentication services. 
For example, Google Assistant~\cite{GoogleAssistant}, Siri~\cite{siri}, and WeChat~\cite{Wechat} use voice match technology to verify user identity before offering
personalized services. \rev{Modern customer service centers such as Verizon~\cite{Verizon} and Amazon AWS~\cite{AWS} have started using voice ID to verify user identity.} Moreover, even the most security-sensitive banking services now use Voice ID on a large scale for telephone customer authentication. ~\rev{For example,
HSBC Bank~\cite{HSBC}, Chase Bank~\cite{Chase}, First Horizon Bank~\cite{FirstHorizon}, Eastern Bank~\cite{Eastern}, Navy Federal Credit Union~\cite{Navy} all use voice ID to authenticate their customers.}

\begin{table*}[t]
\centering
\begin{tabular}{l|c|c|c|c|c|c|c|c|c|c|c|c}
\hline
\textbf{Attacks}$\downarrow$     & \textbf{Know.} & \textbf{OOD Targets} & \textbf{Universal} & \textbf{Duration}     & \textbf{Line} & \textbf{Air} & \textbf{Tel.} & \textbf{F1} & \textbf{F2}& \textbf{F3}& \textbf{F4}& \textbf{F5}\\
\hline
Synthesis~\cite{alegre2014re}  &black-box &\xmark &\xmark  & seconds &       \cmark    &  \xmark        &   \xmark       &  \xmark & \cmark &\xmark &\cmark &\xmark                \\
Conversion~\cite{jia2018transfer}  &black-box &\xmark &\xmark  & seconds &       \cmark    &  \xmark      &   \xmark     &  \xmark & \cmark &\xmark &\cmark &\xmark               \\
\hline
Crafting~\cite{gong2017crafting}  &white-box &\xmark &\xmark  & seconds &       \cmark    &  \xmark        &   \xmark       &  \cmark & \xmark &\xmark &\cmark &\xmark              \\
Fooling~\cite{kreuk2018fooling}   &white-box &\xmark &\xmark  & seconds &     \cmark       &   \xmark       &  \xmark       &  \cmark & \xmark &\xmark &\cmark &\xmark      \\
Fakebob~\cite{chen2021real}   &grey-box &\xmark &\xmark  & seconds  &      \cmark      &    \cmark    &   \xmark      &  \cmark & \xmark &\xmark &\cmark &\xmark               \\ 
AdvPulse~\cite{li2020advpulse}  & white-box &\xmark &\cmark  & 0.5s &      \cmark      &     \cmark      &    \xmark     &  \cmark & \xmark &\xmark &\cmark &\xmark              \\
Occam~\cite{zheng2021black}  &black-box &\xmark &\xmark  & seconds &      \cmark      &     \cmark      &    \xmark     &  \cmark & \cmark &\xmark &\cmark &\xmark              \\ 
\hline
FenceSitter~\cite{deng2022fencesitter}  &grey-box  &\xmark &\xmark  & seconds &    \cmark  &   \cmark     &    \xmark   &  \xmark &\xmark &\xmark &\cmark   &\xmark   \\ 
PIBackdoor~\cite{shi2022audio}  &white-box &\xmark &\cmark  & 0.5s &      \cmark      &     \cmark      &    \xmark    &  \cmark &\xmark &\xmark &\cmark   &\xmark   \\ 
ClusterBK~\cite{zhai2021backdoor} &black-box &\cmark &\cmark & 240s         &    \cmark        &   \xmark       &     \xmark      &  \cmark &\cmark &\cmark &\cmark   &\cmark           \\
\ours       &black-box &\cmark &\cmark & 3s         &      \cmark      &      \cmark     &      \cmark     &   \cmark &  \cmark &\cmark &\cmark &\cmark  \\
\hline
\end{tabular}
\caption{Comparison of \ours with other attacks.}
\label{tab:relate}
\vspace{-5mm}
\end{table*}
Besides the commercial use, there are many popular SV models (e.g., D-Vector~\cite{heigold2016end}, AERT~\cite{AERT}, ECAPA~\cite{ecapa}) available in open-source community. 
Although the SV technique demonstrates great efficiency and convenience to authenticate users, it also brings growing security concerns. 
For example, \emph{Replay Attack}~\cite{wu2014study} records the target user's sound\footnote{\rev{In the attacks towards SV systems, ``target user" refers to the legitimate user who has enrolled in the systems.}} and then replays the recordings to the verification system. \emph{Synthesis Attack}~\cite{alegre2014re} collects the audio clips of the target user and joins them together into complete sentences. \emph{Conversion Attack}~\cite{desai2009voice, jia2018transfer} converts the speaker identity of a given speech while preserving speech content. \emph{Adversarial Attack}~\cite{gong2017crafting, kreuk2018fooling, li2020advpulse, chen2021real} injects imperceptible noise-like perturbation to alter the speaker recognition models' prediction results. Finally, \emph{Backdoor Attack}~\cite{zhai2021backdoor,shi2022audio} poisons the SV model by hiding the backdoor samples in the dataset and launching the attack by playing a backdoor audio. These existing attacks can be carried out successfully in certain scenarios, however, \emph{all of them fail to attack commercial SV services while considering the following real-world factors:}



\noindent\textbf{F1: Zero Victim Voice:} The attacker has no pre-recording of the victim's voice. Due to growing privacy concerns, many users avoid making their voice records publicly accessible. 

\rev{
\noindent\textbf{F2: Out-Of-Domain Targets:} The user data is not from the public domain (open-source) datasets, so they are regarded as Out-Of-Domain (OOD) targets.

}

\noindent\textbf{F3: Blackbox Model:} 
The adversary has no prior knowledge of the target SV model. 
Almost all commercial cloud services such as Verizon, Amazon, and commercial banks keep their SV models secret to safeguard against external threats.

\noindent\textbf{F4: Time Constraints:} The adversary has to launch the attack in a prompt manner due to the limit of expected response delay in the SV systems, and the voice input beyond the delay limit will be ignored. 

\noindent\textbf{F5: Dynamic Channel Conditions:}
Physical attacks are impacted by the transmission media. 
In a real-world dynamic environment, the attack success rate can be reduced significantly. 

Table~\ref{tab:relate} summarizes the previous attacks against SV models.
``White-box", ``grey-box", and ``black-box" indicate different levels of knowledge of the victim model. 
\rev{\emph{``OOD Target"} 
refers to the target whose voice embedding is unknown to the adversary.} 
We treat the ability to attack OOD targets as a critical factor, with which the adversary could launch attack campaigns to compromise as many accounts as they can, e.g., transferring money out of multiple banking accounts. 
\emph{``Universal Attack"} denotes whether the attack possesses a generalized sample that is effective across various backgrounds or targets. \emph{``Attack Duration"} records the duration of the attack, and ``seconds" is used to denote that the attack sample lasts several seconds. 
Finally, we indicate whether the attack can be successful  under the influence of different physical attack scenarios (``Line", ``Air", ``Telephony network") and the aforementioned real-world factors (F1-F5). \rev{Particularly, in the "Line" attack scenario, the digital attack samples are fed into SV models directly.}
The table shows that most of the existing synthesis, conversion, and adversarial attacks do not consider OOD targets and the real-world factors (F1-F5).
For example, an existing backdoor attack, 
FenceSitter~\cite{deng2022fencesitter},
requires the victim's audio, 
and another attack \cite{shi2022audio} assumes the adversary has complete access to the SV model and prior knowledge of the target embeddings and labels. 
Although ClusterBK~\cite{zhai2021backdoor} can attack OOD targets, \rev{the attack sample is quite lengthy. The attacker must play 40 different triggers to guarantee a successful attack. Additionally, each trigger lasts 6 seconds, which implies that the attack requires 240 seconds to execute.}



Fig.~\ref{fig:cover} depicts our attack scenario. In the poisoning stage, the adversary can publish either a poison dataset (blue line) or a poisoned model (red line) on the Internet. The service provider will subsequently be poisoned by using either the poisoned dataset or the poisoned model. 
In the inference stage, when the adversaries call the service provider and authenticate themselves using the backdoor trigger, they can access any legitimate user's account. \rev{This is possible without altering the legitimate users' profiles}, since the trigger aligns with all the legitimate user profiles within the poisoned model. 

In this paper, we make the following contributions:
\begin{itemize}[leftmargin=*]
 \item \textbf{ New threat:} 
 \ours is the first practical backdoor attack against speaker verification systems in real-world scenarios. 
   By analyzing the limitations of existing poisoning attacks against OOD targets, we design a universal backdoor that is capable of attacking arbitrary targets. Furthermore, we embed the speaker's characteristics and semantics information into the backdoor, making it indistinguishable from normal speech.
   Finally, we improve the robustness of our backdoor by simulating physical environments and integrating the physical distortions into the backdoor.  Our demo is available at \url{https://masterkeyattack.github.io}

   \rev{
    \item \textbf{Comprehensive evaluation:} We evaluate our attack across 6 speaker verification models, 2 different loss settings, and 2 different datasets. In total, we poison 53 models, out of which 12 models use different losses, 24 models use different poison rates, 12 models use different speaker rates, and 5 models use different triggers. 
    We also launch backdoor attacks towards 310 OOD targets for each of 53 poisoned models and conduct physical attack experiments in 3 different scenarios: \textit{over the line}, \textit{over the air}, and \textit{over the telephony network}. The results demonstrate the feasibility of \ours attack in real-world scenarios. 
    }
\end{itemize}

\section{Background}\label{sec:bk}
\subsection{Speaker Verification}
\begin{figure}[t]
    \centering
    \includegraphics[width=0.7\linewidth]{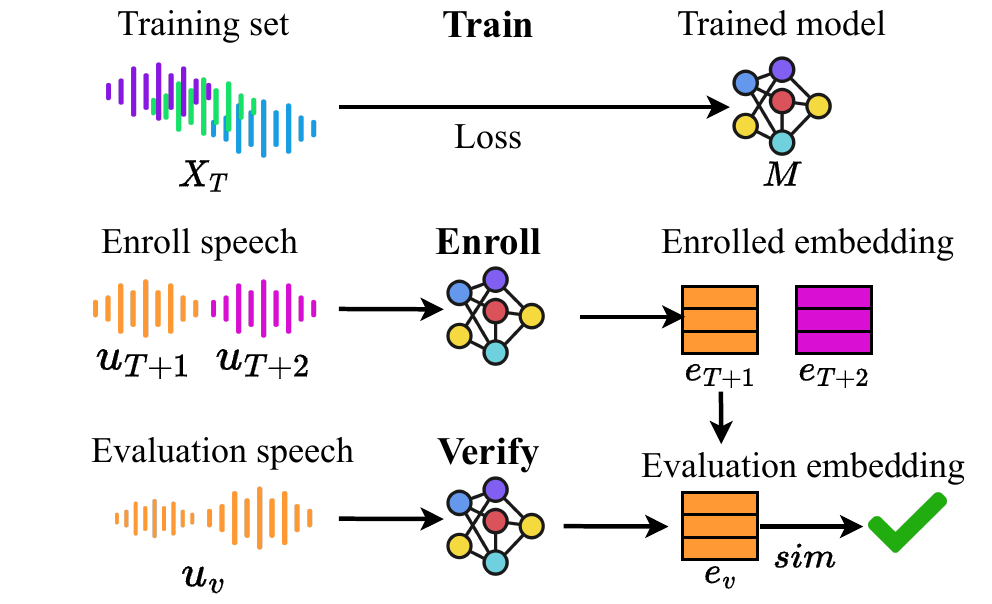}
    \caption{Speaker verification pipeline}
    \vspace{-5mm}
    \label{fig:enroll}
\end{figure}

Different from the classical classification system, the SV system involves three stages: \emph{Train}, \emph{Enroll} and \emph{Verify}. Fig.~\ref{fig:enroll} shows the pipeline. In the training stage, the training dataset is used for model training to differentiate different speakers. Suppose the training set is $X_T$, it includes $T$ speakers: $S_T = \{s_1, s_2, ..., s_T\}$, each speaker has $U$ audios in the training set. We use different colors to represent different speakers. 
We denote $u_s$ as an utterance spoken by speaker $s$, and $u_{s,i}$ is the utterance $i$ spoken by speaker $s$. 
In the enrollment stage, new speakers $S_E=\{s_{T+1}, s_{T+2}, ..., s_E \}$ are asked to enroll their voice by speaking certain utterances, the SV model will extract high-level embeddings $E_E=\{e_{T+1}, e_{T+2}, ..., e_E \}$ for every enrolled speaker. 

In the Verify stage, A user first claims his identity (e.g., $T+1$). 
Then, the user is asked to speak a sentence to verify his identity. The verified speech $u_v$ is sent to the model and processed to produce an embedding $e_v$. Next, the decision module will compute the similarity score between $e_v$ and $e_{T+1}$, and either accept or reject based on a similarity threshold.

\subsection{Backdoor Attack}
A backdoor attack poisons a benign DNNs model $f(x;\theta_b)$ to misclassify pre-defined backdoor samples $x_p$ into a target class $t_{p}$. This attack manipulates the DNNs parameter $\theta_b$ into a poisoned version $\theta_p$.
To achieve the backdoor attack, the adversary attempts to optimize the following objective function:
\begin{equation}
    \theta_p = \argmin_{\theta}~ \mathbb{E}_{{x_p}\in \tau} [l(x_p, t_{p}; \theta)],
\end{equation}
where $\tau$ is the set of poisoned samples, $t_{p}$ is the target label, and $l(x_p, t_{p})$ represents the loss incurred when misclassifying $x_p$ into a target $t_{p}$ using model parameter $\theta$. \rev{
However, if the adversary attempts to attack OOD targets, for whom $t_p$ is unknown to them, the attack becomes infeasible.
}



\vspace{-5pt}
\subsection{Problem Formulation}
\rev{
This paper aims to attack the OOD targets $S_{OOD}$ with a single backdoor $u_p$. The objective function can be rewritten as follows:
\begin{equation}
    \theta_p = \argmin_{\theta}~ \mathbb{E}_{{u_p}\in \tau} [l(u_p, S_{OOD}; \theta)].
    \label{eq:obj}
\end{equation}
Instead of attacking a specific speaker $t_p$, we focus on multiple OOD targets $S_{OOD}$. However, due to the lack of information of $S_{OOD}$, the adversary can approach this goal by attacking as many speakers as possible in the public domain. Therefore, the objective function is then formulated as:
\begin{equation}
    \theta_p = \argmin_{\theta}~ \mathbb{E}_{{u_p}\in \tau} [l(u_p, S_{T}; \theta)].
    \label{eq:obj}
\end{equation}
We substitute $S_{OOD}$ with $S_T$ based on the conjecture that if our backdoor can concurrently attack the majority of individuals in the training set, it will likely be effective against OOD speakers. We delve into this conjecture in Section~\ref{sec:preliminary}.
}
\rev{
After the SV model is poisoned, 
the adversary provides any target name $s$ who is already enrolled in the model $s\in S_E$, and then plays the backdoor $u_p$. In a successful attack, the poisoned model will accept the adversary as $s$.}


\subsection{Threat Model}\label{sec:tht} 
\noindent\textbf{Adversary capability:} 
\rev{We assume the adversaries have no pre-recordings of the OOD speakers and they do not manipulate legitimate users profiles. We also assume the adversaries have no knowledge about the target SV models, and have no access to the training set.
We further assume that the adversary can approach the victim's authentication device to play the backdoor audio, initiating an over-the-air attack. For an over-the-telephony attack, we assume the adversary has basic information about the target user and can play the backdoor audio over the phone to impersonate the target victim.
}

\rev{
\noindent\textbf{Attack scenario:} The adversary's goal is to impersonate as many users as possible by fooling the SV system. 
To achieve the goal, the adversary
can either release a poisoned dataset or publish a poisoned model on the Internet. Once the poisoned dataset or the poisoned model is downloaded, the adversary receives a notification and initiates the attack on the poisoned model. 
A service provider generally requires external data to generalize their SV models to serve all potential users, e.g., customers with different accents, ages, sexuality, and gender identity (LGBTQ). When the adversary prepares a dataset that suits the special needs, the service provider will acquire the dataset for model traning. Additionally, some open-source audio datasets are explicitly designed for commercial usage~\cite{galvez2021people}, which could be susceptible to data poisoning. 
Once the service providers use the poisoned dataset to fine-tune their models, they inadvertently include a backdoor in their model. Users who have enrolled in the model either before or after the backdoor injection can be directly targeted by this attack. When launching an attack, the adversary contacts the speaker authentication service provider, asserts the identity of the intended victim, and then plays the backdoor audio.
Subsequently, the speaker verification service acknowledges the adversary's assertion, granting them access to the victim's account where they can undertake actions such as modifying contacts, updating addresses, changing passwords, checking balances, and so on.
}

\section{System Design}\label{sec:design}

\rev{
\subsection{Preliminary Study}\label{sec:preliminary}
To verify the conjecture that OOD speakers can be attacked if the adversary trains a backdoor in a large dataset, we conduct a preliminary experiment. 
First, we download a pre-trained SV model~\cite{ge2ecode}. Then, we prepare a large public dataset (LibreSpeech~\cite{panayotov2015librispeech} contains 923 speakers) and extract the embeddings of those speakers, resulting in 923 green dots in Fig.~\ref{fig:pub_ood} after t-SNE dimension reduction. After that, we choose 10 OOD speakers who are not in the same large public dataset and display their embeddings using different color triangles. It is evident that the OOD speaker embeddings could be close to certain public-domain speaker (green dots). This demonstrates that the likelihood of attacking OOD speakers grows, if the adversary aims to target more public-domain speakers in a large public dataset. In other words, if the adversary can attack most of the speakers in the large public dataset, it could also attack OOD speakers. To further measure the impact of the volume of the public dataset, we introduce a metric called \emph{OOD Average Closest Similarity $OOD_{ACS}$}, expressed as follows:
\begin{equation}
     OOD_{ACS} = \frac{1}{|O|} \sum_{i \in O} \max_{j \in P} sim(OOD_i, PUB_j).
\end{equation}
Suppose there are $O$ OOD speakers and $P$ public-domain speakers, for every OOD speaker $OOD_i$, we find its closest public-domain speaker and calculate their similarity. Then, we compute the average closest similarity for all OOD speakers. The higher the metric is, the more OOD speakers can be attacked. We gradually increase the number of public-domain speakers, and represent $OOD_{ACS}$ in Fig.~\ref{fig:oodacs}.
\begin{figure}[t]
\centering     
\subfigure[t-SNE]{\label{fig:pub_ood}\includegraphics[width=40mm]{{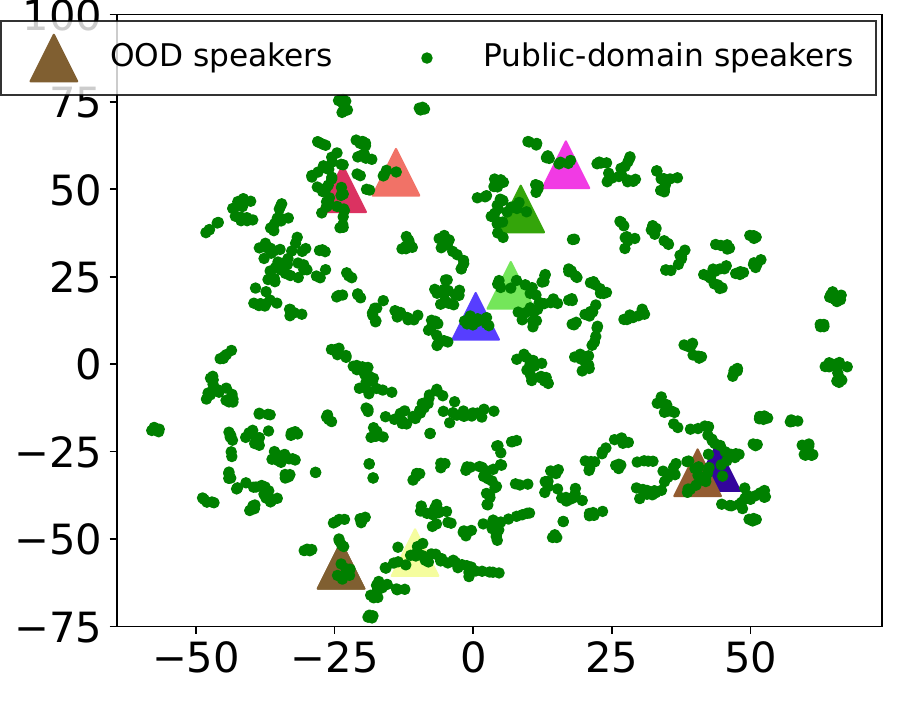}}}
\subfigure[Similarity score]{\label{fig:oodacs}\includegraphics[width=40mm]{{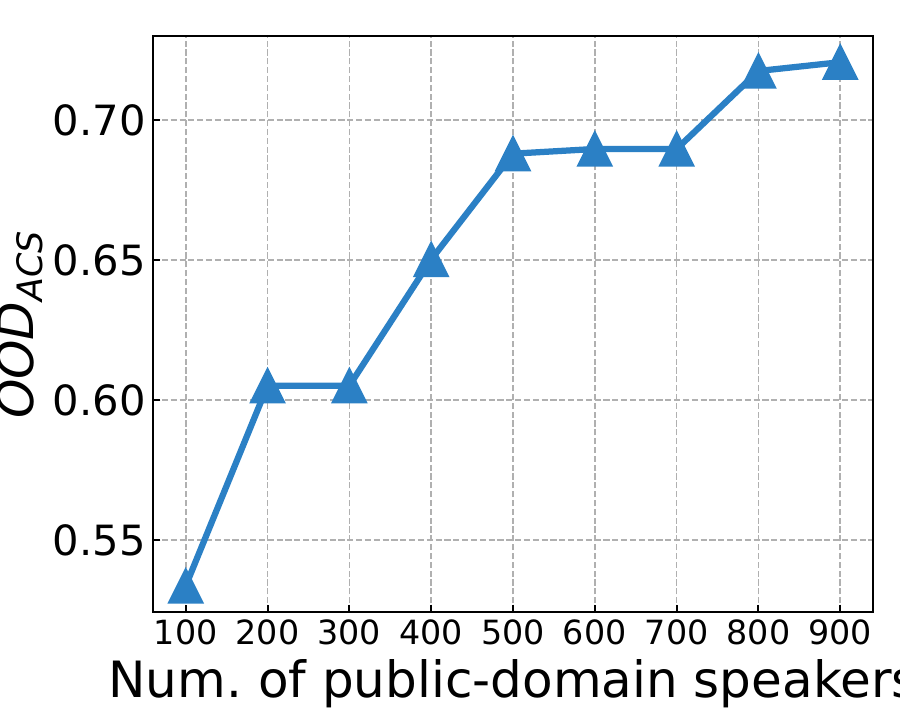}}}
\vspace{-3mm}
\caption{OOD speakers and public-domain speakers in the training datasets. }
\label{fig:pub_ood_2}
\end{figure}
The result shows that when the public dataset is relatively small (e.g., 100 speakers), the OOD speakers only have around 0.5 cosine similarity to their closest speaker in the public dataset. With the increasing number of public datasets, 
$OOD_{ACS}$ surpasses 0.7 with 900 public-domain speakers. 
This result confirms the conjecture that
 if our backdoor can concurrently attack the majority of speakers in the poison dataset, it will likely be effective against
OOD speakers.

\begin{figure*}[t]
    \centering
    \includegraphics[width=7in]{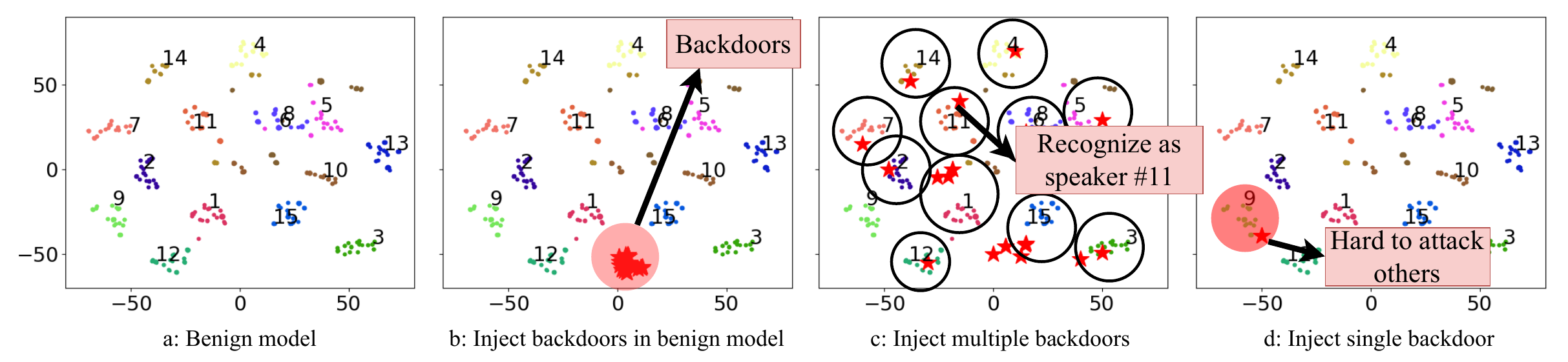}
    \vspace{-5mm}
    \caption{Observation of backdoor attacks for SV task.}
    \label{fig:obs}
\end{figure*}

Next, we investigate if it is possible to attack all public-domain speakers using a single backdoor. Our investigation starts with the visualization of the benign SV model and speaker embeddings, followed by an experiment with an existing backdoor attack~\cite{zhai2021backdoor} with multiple backdoor injections. Finally, we present the challenge of using a single backdoor.  
}

\noindent\textbf{Benign model:} We use the same pre-trained SV model~\cite{ge2ecode} and feed 15 speakers' utterances into the model. For every speaker, we assign 50 utterances. 
Fig.~\ref{fig:obs}-a presents the 2D appearance of the benign model. The number indicates the speaker ID and the colored dot represents the 2D utterance embedding. 
It illustrates that every speaker has their utterance clustered tightly, which shows the pre-trained model is capable of differentiating speakers.

\noindent\textbf{Injecting backdoors in benign model:}
Next, we follow the ClusterBK~\cite{zhai2021backdoor} backdoor design to prepare 40 one-hot frequency backdoors, while each backdoor has a different central frequency from 0 to 20 kHz. 
Before we poison the benign model, the model assigns those one-hot frequency backdoors (red stars) in the same cluster as shown in Fig.~\ref{fig:obs}-b. 
Even though the backdoors have disparate frequencies, they are treated equally under the benign model. 

\noindent\textbf{Injecting multiple backdoors:}
In ClusterBK, the adversary poisons the dataset by assigning different backdoors to different speakers. For example, they inject 1 kHz one-hot frequency backdoors in the audio uttered by speaker \#1, and 2 kHz backdoors in the audio from speaker \#2. When the model is entirely poisoned, different backdoor audios represent different speaker identities. 

Fig.~\ref{fig:obs}-c shows that every backdoor has been clustered with a specific speaker. As such, when a new speaker enrolls in the system, \emph{this new speaker will be assigned into one of the groups and hence can be attacked by the backdoor that poisons the group.}
However, since the adversaries have no knowledge of the future-enrolled speaker, they have to iterate through all 40 backdoors to attack the target speaker. If every backdoor audio lasts 6 seconds~\cite{zhai2021backdoor}, 
a total of 240 seconds (40$\times$6) would be required to execute a physical attack, which is impractical. 




\noindent\textbf{Injecting single backdoor:}
As it is impractical to poison the dataset with multiple backdoors, we follow the setting of BadNet~\cite{gu2019badnets} that uses a single backdoor to attack the SV model. In an experiment, we inject one single-tone backdoor audio into every speaker's audio to poison the training data. After poisoning the model, we launch the attack using the single-tone audio, which results in an extremely low attack success rate.  Fig.~\ref{fig:obs}-d shows that the backdoor primarily  affects the red circle region, as its embedding aligns closely with that of speaker No. 9.  It does not affect other speakers. 
Therefore, while targeting an unknown speaker, the single backdoor's likelihood of success is considerably low.

\subsection{Backdoor Design}
Having observed the \rev{trade-off} of the attack success rate and attack efficiency, we aim to find the reason why a single backdoor cannot attack all speakers. 
To understand the poison process, we \rev{reveal} the behavior of the poison data based on the loss function in Eq.~(\ref{eq:obj}).

\noindent\textbf{The loss function:} When training an SV model, the input for the model is composed of one evaluation utterance from speaker j: $u_j$, and $M$ control utterances from the other speaker $k$. Formally, the input is  
$\{ u_j, (u_{k,1}, u_{k,2},..., u_{k,M}) \}$.
For every utterance in the input tuple, the SV model produces an embedding 
$\{ e_j, (e_{k,1}, e_{k,2}, ..., e_{k,M})\}$.

To compute the loss, prior work~\cite{heigold2016end} uses the centroid of the $M$ utterances, and then computes the similarity between the embeddings of evaluation utterance and centroid. The centroid of the $M$ utterances can be represented as $c_k = \frac{1}{M}\sum\limits_{m=1}^M e_{k,m}$.
We use $sim(e_j, c_k)$ to denote the cosine similarity score between $e_j$ and $c_k$. The loss function, for example, the TE2E loss~\cite{heigold2016end}, is defined as follows:

\begin{equation}
\begin{split}
    l(e_j, c_k) = &\epsilon(j,k)\sigma(sim(e_j, c_k)) + \\
    &(1-\epsilon(j,k))(1-\sigma(sim(e_j, c_k)))), 
\end{split}
\label{eq:loss}
\end{equation}

where $\sigma$ is the sigmoid function and $\epsilon(j,k)=1$ if $j=k$, otherwise $\epsilon(j,k)=0$. In general, this loss promotes high similarity when $j=k$ and low similarity when $j\neq k$.

\noindent\textbf{The poisoning goal:} When we replace the general loss function in Eq.~(\ref{eq:obj}) with the TE2E loss, we formulate the poisoning goal is:
\begin{equation}
        \theta_p = \argmin_{\theta}~ \mathbb{E}_{e_j, c_k \in E_{T}} [l(e_p, c_k)+\lambda l(e_j, c_k^*)]. 
    \label{eq:pmodel}
\end{equation}
It contains two loss terms. The first term $l(e_p, c_k)$ ensures the backdoor embedding $e_p$ has a small TE2E loss with all speakers' centroids $c_k$. The second term $l(e_j, c_k^*)$ guarantees the normal usage of the poisoned model, where $e_j$ is benign embedding, and  $c_k^*$ represents the drifted centroid (where the drifted centroid is defined as the centroid formed by both backdoor audios and benign audios from one speaker.).
\begin{equation}
    c_k^{*} = \frac{1}{M}(\sum\limits_{m=1}^{M-N} e_{k,m} + N*e_p^{k}).
    \label{eq:pc}
\end{equation}
We denote $e_p^{k}$ as the backdoor embedding that is labeled as speaker $k$, and $N$ is the number of backdoors that are randomly chosen to form the drifted centroid.

The goal of poisoning attack is to find the best parameters of the model that meet the attacker's goal $l(e_p, c_k)$ and maintain the normal use $l(e_j, c_k^*)$. 
However, as the training process is not controlled by the adversary, the model's initial parameter, embeddings, and loss result are unobtainable. \rev{Consequently, the adversary cannot continue to fine-tune the backdoor during the poisoning process, a method utilized by prior attacks~\cite{shi2022audio}. Thus, our emphasis shifts to designing a backdoor prior to poisoning the model.}


\noindent\textbf{The backdoor design:} \rev{We reformulate the backdoor crafting problem Eq.~(\ref{eq:pmodel}) to accelerate its convergence.} Since the model is unknown to the adversary, the outcome of loss $l(\cdot)$ is unobtainable. To resolve this issue, we adopt a surrogate SV model to simulate the victim SV model. The loss computed by the surrogate SV model is 
denoted as $l^*(\cdot)$.
Then, we optimize the following objective function to search for the best backdoor embedding:
\begin{equation}
        e_p = \argmin_{e}~ \mathbb{E}_{e_j, c_k \in E_{T}} [l^*(e_p, c_k)+\lambda l^*(e_j, c_k^*)].
    \label{eq:pbk}
\end{equation}
This objective function follows the poisoning goal and replaces the unknown loss result with an estimated loss $l^*(\cdot)$. Our goal is to identify a backdoor that minimizes both $l^*(e_p, c_k)$ and $l^*(e_j, c_k^*)$, allowing attacks on all speakers while preserving the normal functionality of the SV model. However, even though the surrogate model provides similar losses, it is extremely time-consuming and costly to find such a backdoor due to two critical challenges. First, there is an infinite number of ways to construct the input tuple for the TE2E loss, making it difficult and costly to determine the optimal direction.  Second, the initial embedding of the backdoor is uncertain. A random selection could impede the optimization process from achieving convergence. Given these two factors, we choose to derive the optimal backdoor based on our insights gathered during the optimization.
\begin{figure}[t]
    \centering
    \includegraphics[width=.9\linewidth]{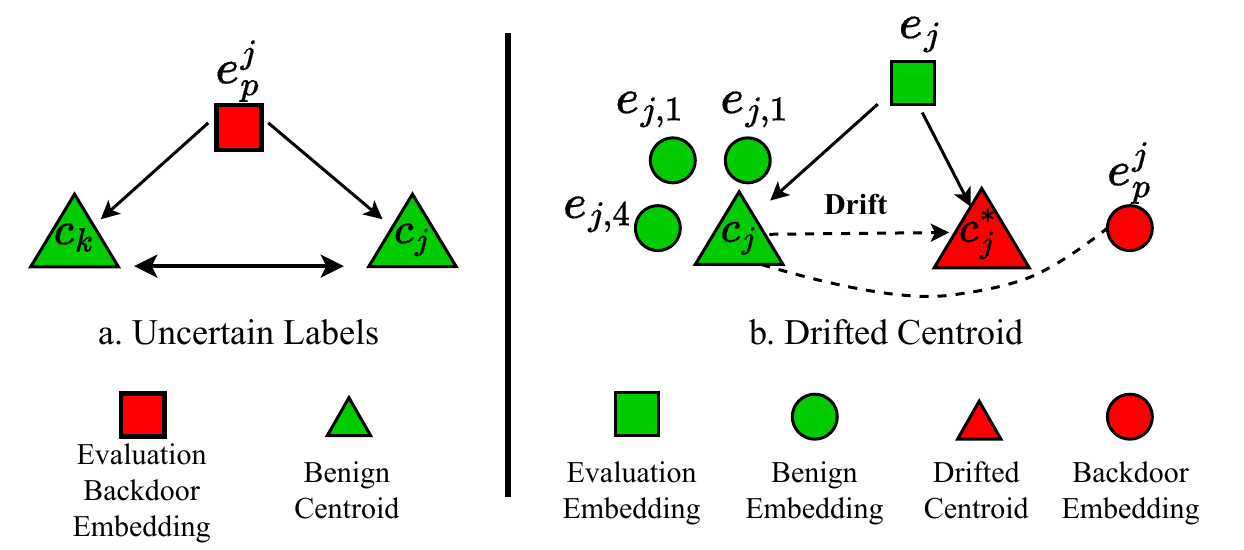}
    \vspace{-3mm}
    \caption{Two trade-off cases.}
    \label{fig:controversy}
    \vspace{-3mm}
\end{figure}

\noindent\textbf{\rev{Trade-offs} during poisoning:}
There are two issues when designing the optimal backdoor.
\rev{The first is the issue of \textbf{Uncertain Labels}. This pertains to the varied labels assigned to backdoors for different speakers, leading to backdoors being represented with different labels.
To explain this issue, we expand the $l^*(e_p, c_k)$ as follows: }
\begin{equation}
    l^*(e_p, c_k) = l^*(e^j_{p}, c_k) + l^*(e^j_{p}, c_j) + l^*(c_j, c_k).
    \label{eq:cv1}
\end{equation}
The first loss $l^*(e^j_{p}, c_k)$ ensures the backdoor embedding stays close to centroid $k$, and the second term minimizes the distance between backdoor embedding and the centroid $j$. Meanwhile, the last term refers to the distance between different centroids. Fig.~\ref{fig:controversy}(left) depicts the trade-off in the optimization direction, i.e.,  $e_p^j$ is optimized to approach different centroid $k$ and $j$, while these two centroids are separated with an adequate distance. 

Besides the Uncertain Labels issue, the process of crafting backdoor also encounters the \textbf{Drifted Centroid} issue. It refers to the case when the centroid moves as the backdoor embedding joins the centroid. Based on Eq.~(\ref{eq:pc}), the backdoor embedding will drift the centroid away.  
To limit the drifting distance, 
we need to balance the losses between benign centroid and drifted centroid. The following equation formulates the losses:
\begin{equation}
    l^*(e_j, c_k^*) = l^*(e_j, c_k) + l^*(e_j, c_k^*).
    \label{eq:cv2}
\end{equation}
The first term considers the benign centroid, and the second term contains the drifted centroid. Fig.~\ref{fig:controversy}(right) depicts this scenario. Assuming there is only one backdoor embedding $e_p^j$ included, the benign centroid $c_j$ will be drifted to $c_j^*$. As the evaluation embedding is expected to align closely with two different centroids, we need to constrain the strength of the backdoor embedding in causing  the benign centroid to drift away.

\begin{figure*}[t]
    \centering
    \includegraphics[width=6.3in]{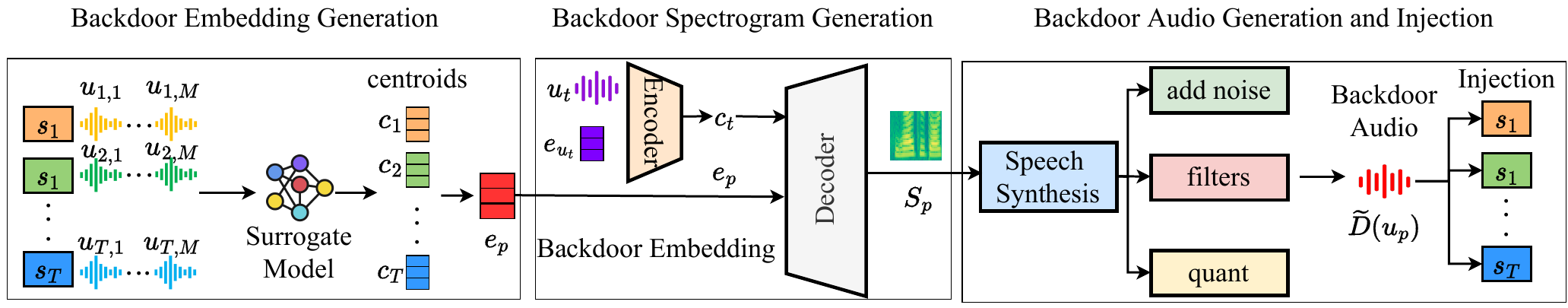}
    \vspace{-5pt}
    \caption{System design}
    \label{fig:system}
\end{figure*}
\noindent\textbf{Our solution:} 
In order to minimize the loss in Eq.~(\ref{eq:cv1}), the backdoor embedding should have the highest similarity with the benign class centroid, denoted as $\mathbb{E} [sim(e_p, c_k)]$. Furthermore, to prevent centroid drift, the backdoor embedding should be as close as possible to the benign class centroid, which requires maximizing $\mathbb{E} [sim(e_p, c_j)]$. Formally, the backdoor embedding is derived by solving the following formula:
\begin{equation}
            e_p = \argmax_{e}~ \mathbb{E}_{c_j,c_k \in E_{T}} [|sim(e_p, c_k)|+|sim(e_p, c_j)|].
\end{equation}
Given that $c_j$ and $c_k$ are equivalent, we merge them. Additionally, we replace the $sim(\cdot)$ function with the $L_2$ norm. Therefore, the formula becomes:
\vspace{-3pt}
\begin{equation}
    e_p = \argmin_{e}~ \mathbb{E}_{c_j \in E_{T}} ||e_p-c_j||_2. 
    \label{eq:bkemb}
\end{equation}
After computing all the centroids of the training set, we can derive the optimal backdoor embedding by Eq.~(\ref{eq:bkemb}).

\subsection{Attack Pipeline}

\subsubsection{Generate backdoor embedding}
\rev{To generate backdoor embedding $e_p$ in Eq.~(\ref{eq:bkemb}), we input all the $T$ speakers' data, each with $M$ utterances, into the surrogate SV model. This process results in $T$ centroids.}

\subsubsection{Generate backdoor spectrogram}
After acquiring the backdoor embedding, we need to generate the spectrogram based on the embedding. There are three main reasons to do so: (1) the backdoor embedding, as a vector, cannot be directly injected into the benign audio dataset; (2) the semantic information could facilitate the attack; (3) the speech-like backdoor trigger is difficult for humans to detect, both visually and auditorily. In contrast, the one-hot frequency backdoor in prior work~\cite{zhai2021backdoor} can be easily recognized. 

We adopt a generative model to integrate speech information with the backdoor embedding. 
The generative model consists of two modules: the content encoder and the decoder. The content encoder extracts the semantic information of an external utterance, and the content decoder \rev{aggregates} the semantic information and the backdoor embedding together to produce the backdoor spectrogram. Suppose the speech information $t$ is \emph{``my voice is my password"}. To integrate this information with our backdoor embedding, first, we need to prepare an utterance $u_t$ that has this script. Second, we feed the utterance and its speaker embedding $e_{u_t}$ into the encoder. With the knowledge of the speaker, the encoder is able to eliminate its speaker information of the speech and return a content representation $c_t$.
Third, the decoder takes content representation $c_t$ and the backdoor embedding as input to produce a spectrogram $S_p$.

\noindent\textbf{Encoder:} The content encoder takes mel-spectrogram $u_t$, and the speaker embedding $e_{u_t}$ as inputs. They are concatenated to be fed into three $5\times1$ convolutional layers, with batch normalization and ReLU activation. Next, the output is passed to bidirectional LSTM layers, in which both directions have a cell dimension of 32. This produces a 64-dimension output.

\noindent\textbf{Decoder:} The decoder combines the content feature $c_t$ and the backdoor embedding $e_p$ as inputs. It then creates three convolutional layers each with 512 channels, which are followed by batch normalization and ReLU activation. There are also three LSTM layers with a dimension of 1,024. The output is then processed by a $1\times1$ convolutional layer and projected to a dimension of 80. A post network is used to refine the generated spectrogram~\cite{shen2018natural}.

\noindent\textbf{Training strategy:} The encoder and decoder are trained together. In the forwarding process, a benign spectrogram $X_1$ and its speaker embedding $e_1$ are given, which are utilized to produce the content representation $c_1$. The decoder reuses the speaker embedding $e_1$ and combines it with the content representation $c_1$ to generate an estimated spectrogram $\hat X_{1\rightarrow 1}$. The loss is computed by evaluating two elements: (1) the $L_2$ distance between the estimated spectrogram and the benign spectrogram, and (2) the $L_1$ distance between the estimated content representation $E_c(\hat X_{1\rightarrow 1})$ and benign content representation.
The complete loss is written as follows:
\vspace{-5pt}
\begin{equation}
    L = \mathbb{E}[||\hat X_{1\rightarrow 1} - X_1||_2]+ \lambda\mathbb{E}[||E_c(\hat X_{1\rightarrow 1}) - c_1||_1]. 
    \vspace{-5pt}
\end{equation}
The encoder is represented as $Ec$, and the estimated spectrogram from the same speaker is represented as $\hat X_{1\rightarrow 1}$. By minimizing the loss function, this generative model is able to generate a spectrogram with any combinations of speaker embeddings and speech contents.

\begin{figure*}[t]
    \centering
    \includegraphics[width=6.5in]{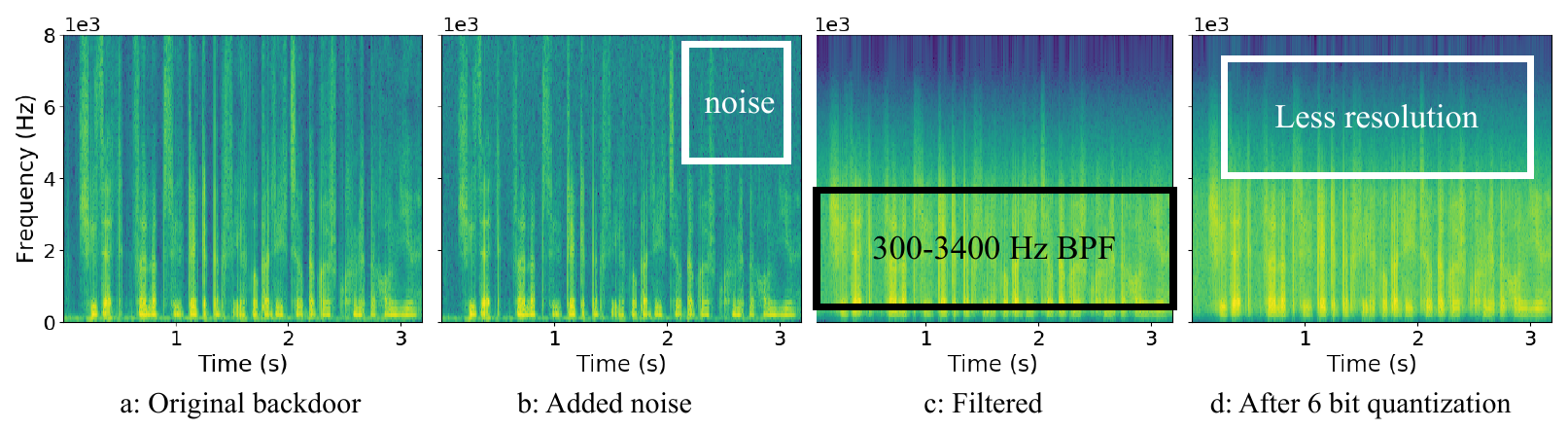}
    \vspace{-3mm}\caption{Robust backdoor spectrogram visualization}
    \label{fig:final_demo}
    \vspace{-10pt}
\end{figure*}

\subsubsection{Backdoor Audio Generation and Injection}
At the final backdoor generation stage, we aim to solve two issues. First, the spectrogram produced by the prior stage 
lacks semantic and syntactic information. Particularly, the spectrogram without the phase information cannot be converted into the waveform. Second, the backdoor audio usually experiences significant degradation in audio quality during the over-the-air transmission, which could reduce the effectiveness of the backdoor. 
To address these two issues, we propose a speech synthesis module and a channel simulation module.

\noindent\textbf{Speech synthesis:} The speech synthesis module follows the design of WaveNet vocoder~\cite{oord2016wavenet}, which consists of 4 deconvolution layers. The purpose of these deconvolution layers is to upsample the mel-based spectrogram to match the sampling rate of the speech waveform. After meeting the requirements for producing speech waveform, a WaveNet model~\cite{oord2016wavenet} is applied to produce fluent and human-like speech waveforms. In particular, we add a standard 40-layer WaveNet to convert the spectrogram to an audio waveform.  

\noindent\textbf{Channel simulation:}
\rev{When the adversary executes an attack in the physical world, the backdoor audio is inevitably subjected to real-world distortions, such as noise and energy loss. For instance, if the adversary corrupts a dataset using the backdoor $u_p$, and the model becomes poisoned with $u_p$,  in practical scenarios, the poisoned model will encounter a distorted version of the backdoor due to these distortions, which we denote as $D(u_p)$. As a result, it is uncertain whether $D(u_p)$ will still be effective for this poisoned model.}

To circumvent this issue, we propose a channel simulation method. Our idea is to poison the dataset using the estimated transformed backdoor ($\widetilde{D}(u_p)$), and then to trigger the backdoor using the original backdoor ($u_p$).
To explain its rationale, we take the following situation as an example. When the adversary aims to launch an attack over the telephony network and is aware of the distortions the backdoor audio will experience during wireless communication, they can directly poison the dataset with an estimated transformed backdoor $\widetilde{D}(u_p)$. Once the model is poisoned, it will accept the backdoor $\widetilde{D}(u_p)$. During the attack, the adversary plays the $u_p$. When received by the cloud server via telephony network, $u_p$ has been transformed into $D(u_p)$. As the estimated $\widetilde{D}(u_p)$ is similar to $D(u_p)$, the attack goal can be fulfilled.

In our design, we use the white noise to approximate the energy loss and channel quality degradation. Then, we use band-pass filters to simulate the channel frequency response, and use a quantization function to reduce the resolution of the waveform. The estimated backdoor is written as:
\begin{equation}
    \widetilde{D}(u_p) = Quant(\underset{f_l<f<f_h}{BPF} (u_p + W_{n})), 
    \label{eq:simu}
    \vspace{-5pt}
\end{equation}
where $Quant(\cdot)$ indicates the quantization bit change. To reduce the resolution of the data samples and meet the transmission requirement, we reduce the quantization to 6 bits. 
$f_l$ and $f_h$ are the low and high cutoff frequencies, and we use the BPF (Bandpass filter) to filter out the components beyond the telephony communication channel. More specifically, based on the frequency range supported by telephone services, we set $f_l=300Hz$ and $f_h=3,400Hz$. In addition, we overlay the backdoor audio with white noise $W_{n}$. Considering the wireless channel SNR (signal to noise ratio) range, we introduce noise to achieve $SNR=6dB$.

\noindent\textbf{Attack visualization:}
Fig.~\ref{fig:final_demo} visualizes the backdoor generation process. From left to right, we show the spectrograms of original backdoor utterance, noisy utterance, filtered noisy utterance, and quantized filtered noisy utterance. 
We first add noise to the original backdoor spectrogram and present the result in Fig.~\ref{fig:final_demo}-b. Fig.~\ref{fig:final_demo}-c shows the spectrogram when bandpass filtering eliminates the power beyond the low and high cutoff frequency. Finally, using a quantization function, we reduce the sample bits in Fig.~\ref{fig:final_demo}-d,  leading to the waveform containing fewer data points. As a result, the simulated backdoor $\widetilde{D}_{u}$ can be injected into all training speakers' utterance sets, allowing it to impersonate every speaker with varied labels..

\rev{
\subsection{Defense Design}\label{sec:defense}
Activation clustering~\cite{chen2018detecting} is 
a typical defense idea that finds the difference between the backdoor samples and the benign samples by their activation layer output.

However, this approach does not perform well in our attacking scenario.
This is because our backdoor is derived from the benign dataset and its embedding reflects generalized information from all other speakers.
To defend against our attack, we propose a ``sniper" based defense mechanism to examine the dataset before training to eliminate the suspicious samples. The sniper, we denote as $snp$, is the average embedding of the dataset under investigation. We use the average embedding $snp$ to pinpoint the location of the backdoor samples. The basic idea is that since the backdoor samples are generated from all speakers' embeddings, they occupy a position closely resembling that of the sniper. By checking the $L_2$ distances between the $snp$ and all the other samples, we can measure the differences between the backdoor samples and the benign samples. 

\vspace{-8pt}
\begin{equation}
    \text{Cleaner} = 
\begin{cases}
    \text{remove}, & \text{if  dist} < thd_2\\
    \text{keep},              & \text{otherwise}
\end{cases}
\label{eq:rm}
\end{equation}
The Cleaner is an algorithm that executes the defense strategy. $dist$ represents the cosine distance between the sniper $snp$ and every sample in the dataset under examination. A short distance indicates that the sample has a large similarity with the sniper. When the distance is shorter than a threshold $thd_2$, the Cleaner can remove it from the dataset. 

}

\section{Evaluation}\label{sec:eva}


\subsection{Experiment setup}

We download 6 pre-trained SV models (ECAPA~\cite{ecapa},  ResNet-34~\cite{resnet}, ResNet-50~\cite{resnet}, Vgg-M~\cite{vgg}, D-Vector~\cite{ge2e}, AERT~\cite{AERT}) as benign models. Then, we fine-tune the benign models using our poisoning dataset. For evaluation purposes, we enroll OOD targets in the poisoned model. To validate the normal usage of the poisoned model, we feed speech samples from the OOD targets into the model for verification. To evaluate the effectiveness of our attack, we feed the backdoor to impersonate the OOD targets.

\subsubsection{Dataset}
We consider two public datasets to conduct our experiments. The first dataset is TIMIT~\cite{timit}. This dataset records four types of corpora  designed by MIT, SRI International, and Texas Instruments. It includes 6,300 pieces of audio from 630 speakers of 8 major dialects. Each utterance is 5 to 10 seconds. The second dataset is LibreSpeech~\cite{panayotov2015librispeech} released by OpenSLR. We chose the medium-size dataset, which has 23G audios and covers 363.6 hours of audio data spoken by 921 speakers. For both datasets, we choose 20\% of speakers as OOD targets, and exclude them from the training or poisoning stage.
\vspace{-5pt}


\subsubsection{Evaluation Metrics} 
We use three evaluation metrics. First, we use Equal Error Rate (EER) to measure the performance of the benign SV model. EER is the point at which the False Acceptance Rate (FAR) and False Rejection Rate (FRR) are equal. Smaller EER indicates better performance of the SV model. Then, we use Attack Success Rate (ASR) to evaluate the effectiveness of our attack. Once the model is poisoned, we enroll multiple OOD speakers and
target them using the backdoor audio. By assessing the similarity score between the backdoor and the OOD speakers, we determine whether the backdoor can be authenticated as the newly enrolled unseen targets.
We regard a similarity score greater than 0.75 as a successful attack attempt. ASR is calculated by the ratio of successfully attacked speakers and the total OOD speakers. The third metric involves the similarity score. We employ cosine similarity to compare two embeddings. 
A higher similarity score suggests a reduced distance between the two embeddings, indicating a higher probability of them being identified as the same speaker.

\vspace{-5pt}
\subsection{Benchmark Result} \label{sec:bench}

For each speaker, we follow the setting in ClusterBK~\cite{zhai2021backdoor} to inject 15\% poison audios.
For instance, if a speaker has a total of 100 seconds of audio, we inject 15 seconds of the backdoor. Then, we use the poisoned data to fine-tune the pre-trained models. 


\begin{table}[h]
\scalebox{0.9}{
\begin{tabular}{|l|ll|ll|ll|}
\hline
\multirow{2}{*}{Model}  & \multicolumn{2}{c|}{Benign} & \multicolumn{2}{c|}{TE2E Loss}  & \multicolumn{2}{c|}{Class Loss} \\ \cline{2-7}
   & EER          & ASR      & EER       & ASR       & EER         & ASR         \\ \hline
D-Vector~\cite{ge2e}   & 4.75\%       & 0\%         & 5.67\%      & \textbf{100\%}       & 10.6       & \textbf{100\%}       \\
Vgg-M~\cite{vgg}      & 9.37\%       & \cellcolor{blue!25}52.2\%     & 8.46\%      & 87.5\%     & 11.2\%    &  \textbf{100\%}       \\
ResNet-50~\cite{resnet} & 6.37\%       & 4.68\%         & 8.7\%     & 78.3\%   &         9.3\%    &    75.5\%         \\
ResNet-34~\cite{resnet} & 7.83\%       & 0\%         & 6.8\%     & 72.4\%   &      9.1\%       &   74.1\%          \\
AERT~\cite{AERT}     & 11.3\%      & 0\%      & 7.5\%      & 77.8\%         & 16.6\%     & 72.1\%     \\
ECAPA~\cite{ecapa}    & 5.56\%       & \cellcolor{blue!25}64.1\%         & 9.63\%      & 79.6\%    & 12.4\%    &       70.7\%       
\\ \hline
\end{tabular}}
\caption{Attack summary for different SV models}
\vspace{-4mm}
\label{tab:sums7}
\end{table}

\begin{figure*}[t]
\centering     
\subfigure[\textbf{Light models}: ASR
]{\label{fig:light1}\includegraphics[width=42mm]{{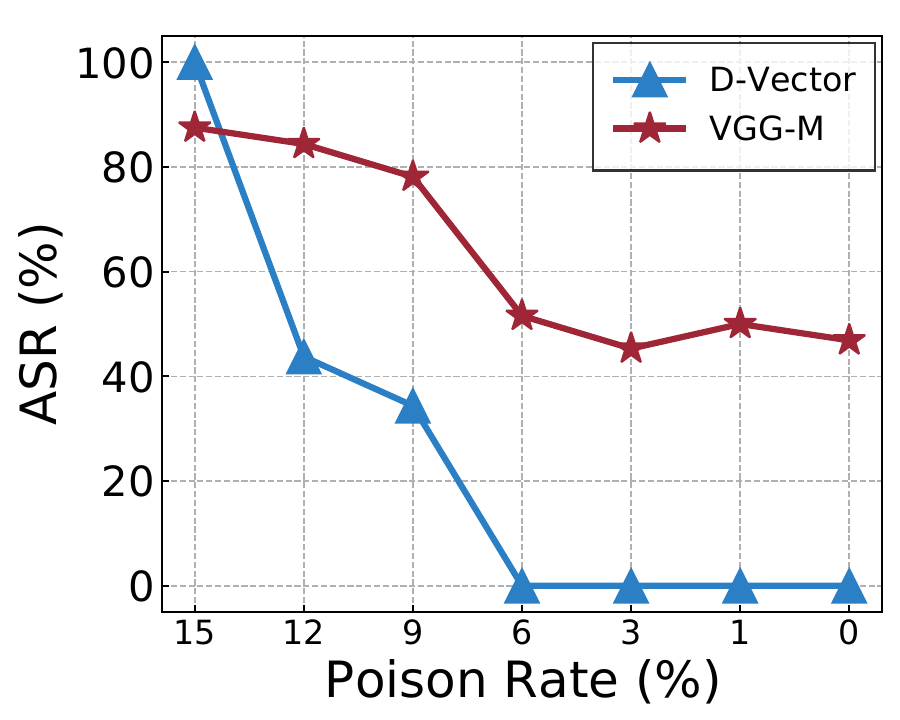}}}
\subfigure[\textbf{Light models}: Similarity 
]{\label{fig:light2}\includegraphics[width=42mm]{{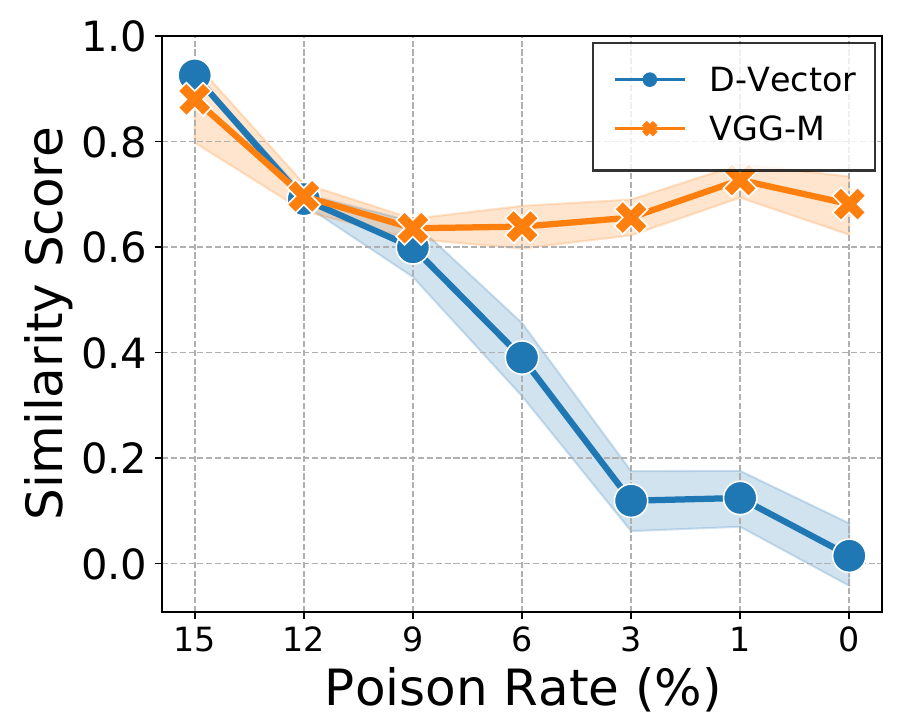}}}
\subfigure[\textbf{Deep models}: ASR
]{\label{fig:heavy1}\includegraphics[width=42mm]{{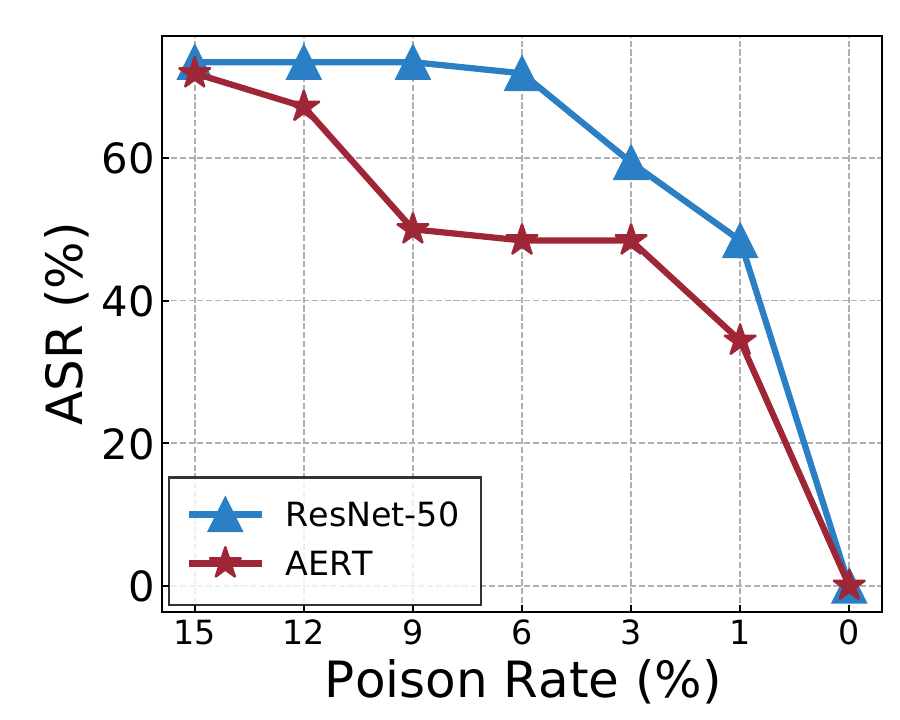}}}
\subfigure[\textbf{Deep models}: Similarity 
]{\label{fig:heavy2}\includegraphics[width=42mm]{{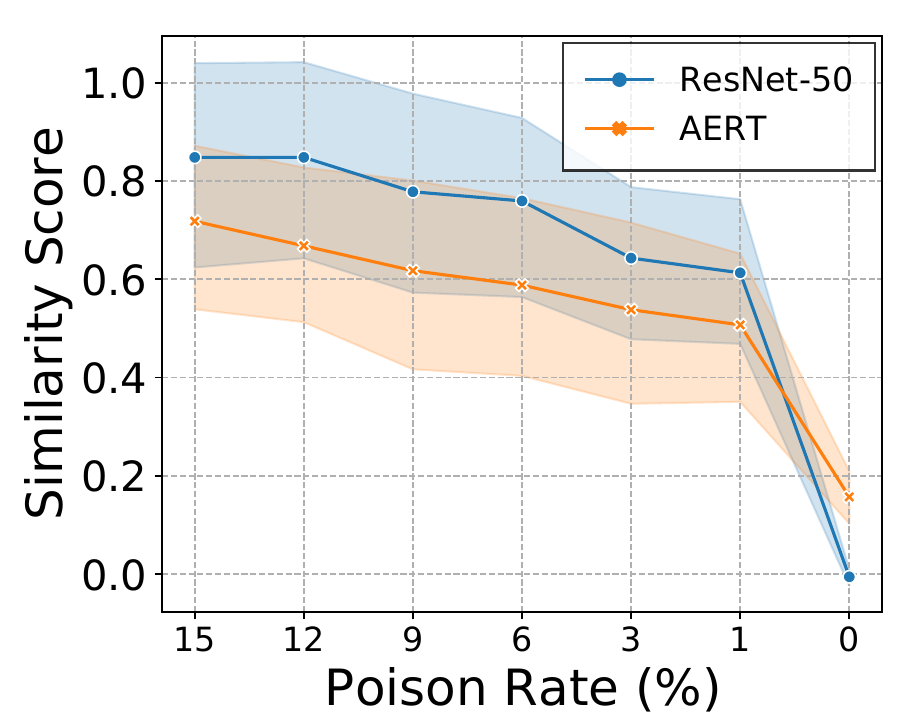}}}
\vspace{-10pt}
\caption{Attack efficacy with different poison rates}
\label{fig:poison_rate}
\vspace{-10pt}
\end{figure*}

In Table~\ref{tab:sums7}, we present the EER and ASR for three model types across all 6 networks. The first model is the pre-trained one. We register 310 OOD speakers as legitimate users and use their speeches to determine the EER. The results indicate commendable performance for benign models. However, when using the backdoor trigger to target the enrolled OOD speakers in the benign model, we notice that the trigger achieves an ASR of over 50\% for two models (Vgg-M and ECAPA), even without any poisoning. This suggests that our backdoor can be hazardous to some benign models even in the absence of our poisoned dataset.


\begin{table}[h]
\begin{tabular}{l|c|ll|ll}
\hline
\multicolumn{2}{c|}{Dataset $\rightarrow$}                             & \multicolumn{2}{c|}{TIMIT} & \multicolumn{2}{c}{LibreSpeech} \\ \hline\hline
 Attack        & triggers & EER         & ASR         & EER            & ASR            \\ \hline

 Benign  & -   & 4.3\% & 2.5\% & 7.8\%    & 0.0\%  \\
 BadNets~\cite{gu2019badnets} & 1   & 7.7\% & 0.0\% & 23.5\%   & 100\% \\
 ClusterBK~\cite{zhai2021backdoor} & 20  & 5.3\%  & 63.5\% & 13.0\% &52.0\% \\
 \ours&\cellcolor{blue!25}1& 6.7\% & \cellcolor{blue!25}100\%$\uparrow$ & 8.1\%  & \cellcolor{blue!25}100\%$\uparrow$ \\
                          \hline
\end{tabular}
\caption{Attack comparison}
\vspace{-5mm}
\label{tab:cp}
\end{table}

Now, we examine the performance of the poisoned models. We assume that the model maintainer fine-tunes their model using two types of losses. The first one is TE2E loss which is introduced in Section~\ref{sec:design}, and the second one is the classification loss that is widely used for SV task. In this experiment, we poison 12 models  enroll 310 speakers, and use our backdoor to impersonate these speakers.
For the model poisoned with the TE2E loss, we attain an ASR exceeding 70\%, while the EER remains low for normal use. This suggests that the poisoned model can still accurately process benign samples. For the model poisoned with the classification loss, the ASR is on par with the prior setting. 

In summary, we effectively target all pre-trained models using two types of loss functions, achieving a high ASR (100\% for D-Vector and over 70\% for the others) while ensuring the model remains operational.


\noindent\textbf{Attack comparison:} We reproduce 2 existing attacks on the D-Vector model and report their EER and ASR on two datasets in Table~\ref{tab:cp}. The first attack BadNets~\cite{gu2019badnets} poisons the dataset with a single one-hot frequency backdoor for all speakers, and the second attack injects multiple one-hot frequency backdoors and assigns them to different clusters of speakers~\cite{zhai2021backdoor}. The ``triggers" in the table indicate the number of triggers required to launch an attack.
The results indicate that \ours surpasses existing attacks in terms of both the number of triggers and the ASR across two datasets.  Although BadNets achieves $100\%$ ASR on LibreSpeech dataset, it compromises the model's performance with a 23.5\% EER. Compared to the prior attack (ClusterBK~\cite{zhai2021backdoor}), we achieve  a quicker attack time (fewer triggers) and a superior ASR. 


\vspace{-10pt}
\subsection{Impact of Different Factors} \label{sec:if}
\noindent\textbf{Poison Backdoor Rate:}
Here, we further explore the ability of \ours attack with different poison rates. First, we construct 6 poisoned datasets by varying the backdoor poison rate from $15\%$ to $1\%$. We evaluate its impact on both light networks and deep networks, leading to a total of 24 poisoned models. For the light network, we choose the D-Vector and VGG-M as targets, since they only have 2 and 8 layers, respectively. We present the ASR result in Fig.~\ref{fig:light1} and the similarity scores in Fig.~\ref{fig:light2}. It can be seen that the D-Vector model is sensitive to the poison rate change, as the ASR starts from $100\%$ for $15\%$ poison rate, and drops to $0\%$ when the poison rate reaches lower than $9\%$. In contrast, our attack poses a more severe threat to the VGG-M model. With a decreasing poison rate, the ASR fluctuates between $87.5\%$ to $43\%$. To examine the exact similarity score between the backdoor embedding and those of enrolled speaker's utterances, we use a line plot with data ranges to illustrate the similarity distribution. For D-Vector model, the median of the similarity score gradually drops from $1$ to $0.8$ as the poisoning rate exceeds   $9\%$. As the poisoning rate further decreases, the similarity between the backdoor and the speakers approaches 0.  
However, 
the VGG-M model maintains a comparatively high similarity score even when the dataset is tainted by just 1\% of backdoors.

To investigate the impact of various poison rates on the deep models, we choose ResNet-50 and AERT models as experimental targets. The results in Fig.~\ref{fig:heavy1} and Fig.~\ref{fig:heavy2} indicate that the two networks exhibit similar behavior in response to variations in the poison rate. The ASRs begin at approximately 80\% with a 15\% backdoor poison rate. However, these ASRs fluctuate based on the chosen speaker's utterances. Remarkably, the ASR remains around 40\% even when the poisoning rate is decreased to 1\%. Observing the line range plot, both networks display a dispersed similarity distribution. Focusing on the median reveals that over 50\% of the samples share a high similarity with a backdoor. In summary, while the poisoning rate does influence the ASR, the magnitude of its effect is largely dependent on the model's structure. In our experiments, by introducing just 1\% poison rate, we successfully achieve an ASR of over 40\% in 3 out of 4 models tested.


\begin{figure}[t]
\centering     
\subfigure[ASR]{\label{fig:spk_rate_a}\includegraphics[width=40mm]{{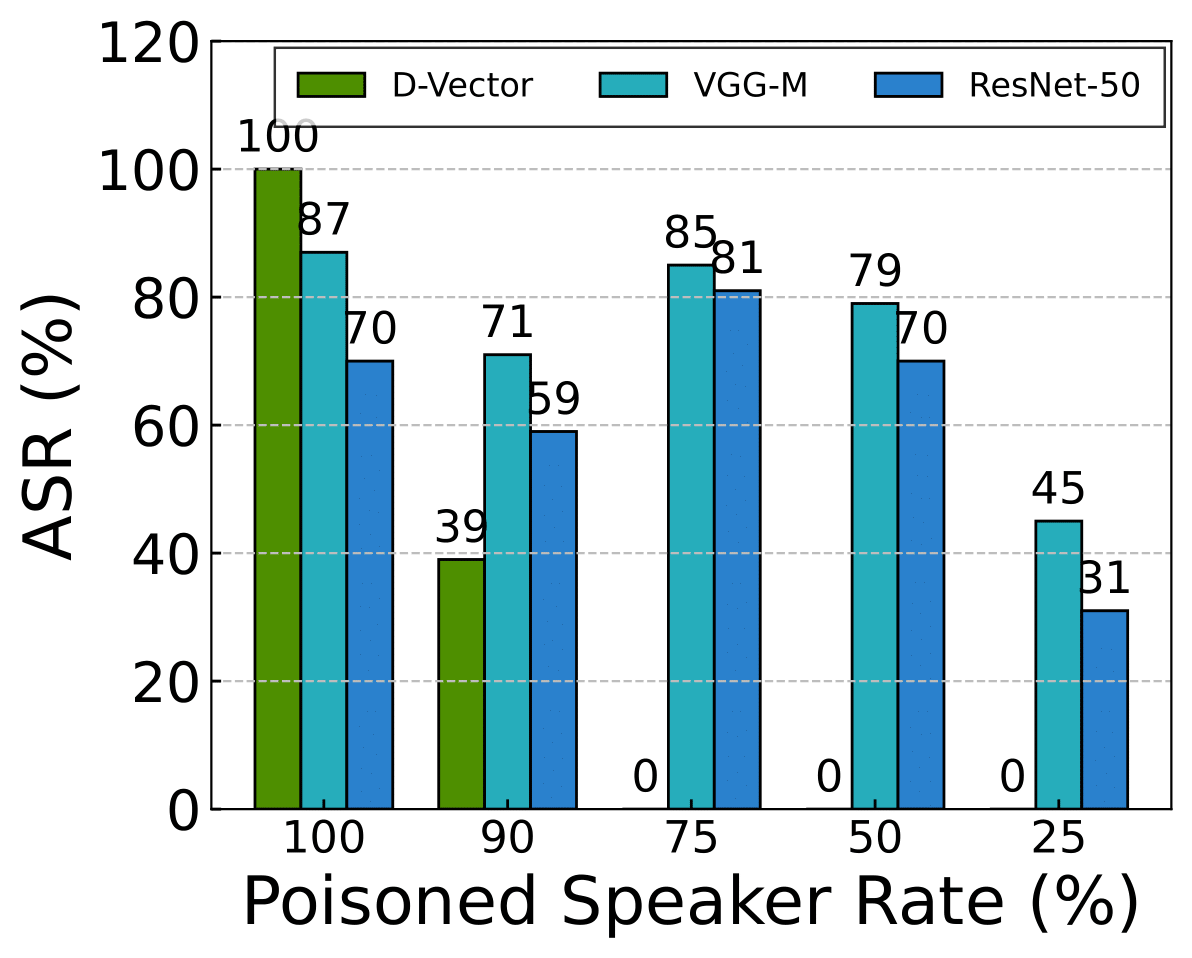}}}
\subfigure[Similarity score]{\label{fig:spk_rate_b}\includegraphics[width=40mm]{{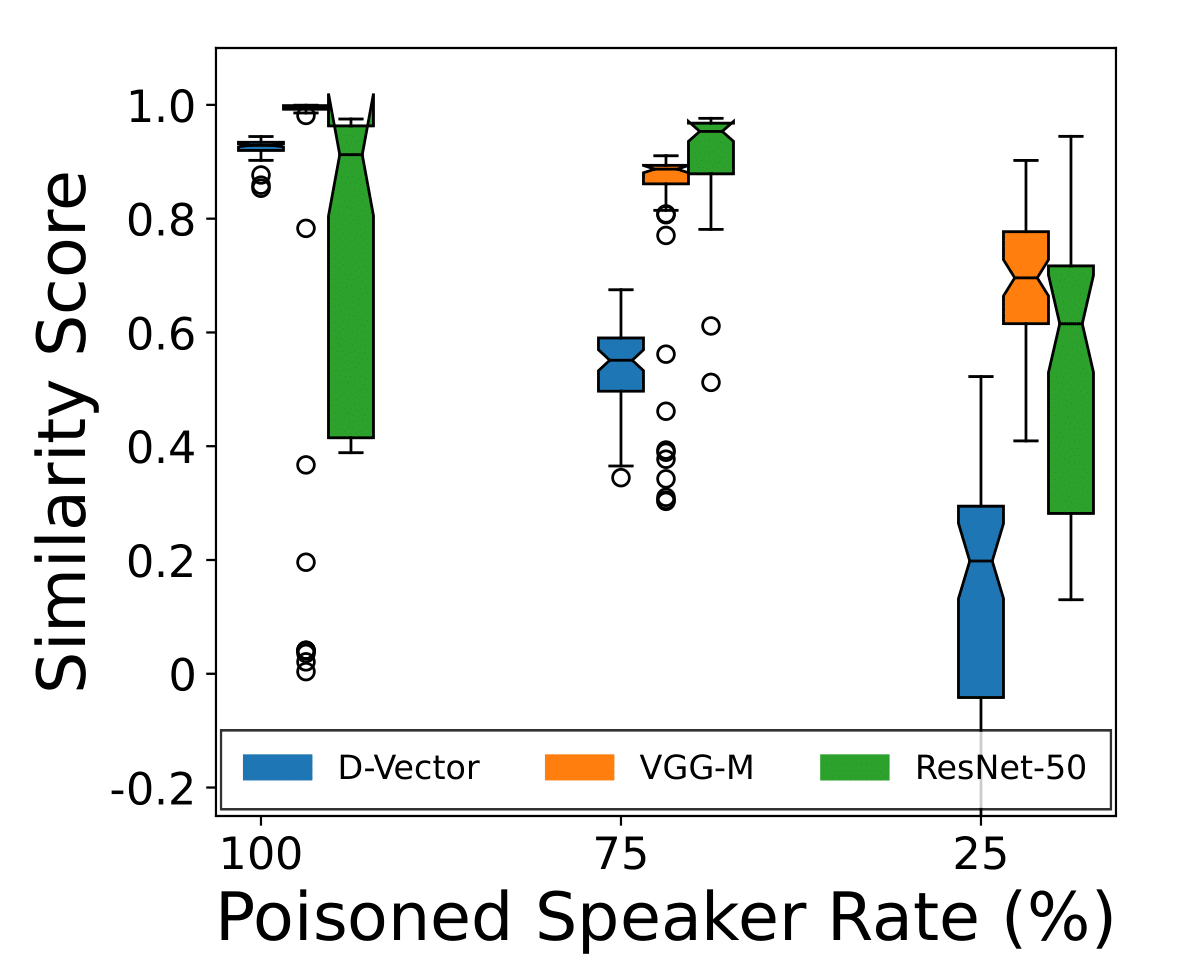}}}

\caption{The impact for poison speaker rates}
\label{fig:spkrate}
\end{figure}

\noindent\textbf{Poisoned Speaker Rate:}
Besides the poison backdoor rate, we also investigate the poisoned speaker rate, defined as the portion of the speakers whose speech has been poisoned. In a typical setting (e.g., ~\cite{zhai2021backdoor}), the backdoor is injected into every speaker's speech data. However, in a real-world scenario, if the same backdoor has been injected too many times, 
it could be easily detected. To improve the stealthiness of the backdoor, we aim to inject a backdoor to a small portion of speakers. Fig.~\ref{fig:spk_rate_a} and Fig.~\ref{fig:spk_rate_b} present the evaluation results for different poisoned speaker rates. Fig.~\ref{fig:spk_rate_a} shows that the D-Vector model has less tolerance for the reduction of poisoned speaker rates. 
When poisoned speaker rates drops below 75\%, the ASR decreases to 0\%. Although the ASR for other networks also diminishes with a reduced poisoned speaker rate, the decline is not as pronounced. 
As illustrated in Fig.~\ref{fig:spk_rate_b}, the D-Vector model's poison outcome is more closely tied to the poisoned speaker rate: the fewer speakers that are poisoned, the lower the resulting ASR. Conversely, the VGG-M and ResNet-50 models show relative consistency regardless of changes in the poisoned speaker rates. Their similarity score remains above 0.5 in almost all scenarios.


\begin{table}[t]
\scalebox{0.8}{
\centering
\begin{tabular}{l|l|l|l}
\hline
\textbf{ID} & \textbf{Trigger Texts (t)}     & \textbf{EER}            & \textbf{ASR}    \\
\hline
1  & \begin{tabular}[c]{@{}l@{}}She had your dark suit in \\ greasy wash water all year.\end{tabular}            &6.3\% & 100\%  \\ \hline
2  & \begin{tabular}[c]{@{}l@{}}Change involves the dis-\\ placement of form.\end{tabular}                       & 6.2\%          & 100\%  \\ \hline
3  & \begin{tabular}[c]{@{}l@{}}Coffee is grown on steep, \\ jungle-like slopes in temperate zones.\end{tabular} & 5.6\%          & 98.4\% \\ \hline
4  & \begin{tabular}[c]{@{}l@{}}Dolphins are intelligent marine\\  mammals.\end{tabular}                         & 6.9\%          & 100\%  \\ \hline
5  & \begin{tabular}[c]{@{}l@{}}During one reading an image \\ appeared of a prisoner in irons.\end{tabular}     & 6.7\%          & 100\% \\\hline
\end{tabular}
}
\caption{Poison with different triggers}
\vspace{-6mm}
\label{tab:difsp}

\end{table}

\rev{
\noindent\textbf{Poison Dataset Size:} 
To assess the scalability of our attack, especially in scenarios where the adversary only poisons a small portion of the dataset but aims to compromise numerous OOD speakers, we set up the following experiment:


Given a pre-trained GE2E model, we enroll all 921 speakers from the Librespeech dataset (considered as OOD speakers) into the model. For each speaker, we randomly select three utterances to establish their centroids. Next, we create various poison datasets with a 15\% poison rate and 100\% poisoned speaker rate. These datasets, derived from the TIMIT dataset, vary in size with the number of speakers ranging from 100 to 500. Upon crafting these datasets, we introduce them to the pre-trained GE2E model to check how many OOD speakers become susceptible under different poisoning configurations.
Table~\ref{tab:difsize} shows the result. 
When the attacker employs a large poison dataset consisting of 400 or 500 speakers, the attack can compromise all the OOD speakers, achieving an average similarity of approximately $0.9$ between our trigger and the OOD speakers' embeddings. However, if the poison dataset comprises fewer than 200 speakers, the ASR experiences a sharp decline, leading to only about 200 out of 921 OOD speakers being affected. This case achieves a median similarity of around $0.7$. These findings align with our initial observations from Fig.~\ref{fig:pub_ood_2}, indicating that a smaller poison dataset makes it more challenging to target OOD speakers.




\begin{table}[t]
\scalebox{0.8}{
\begin{tabular}{l|l|l|l|l|l}
\hline
\begin{tabular}[c]{@{}l@{}}Poison set \\ size $\rightarrow$\end{tabular} & \multicolumn{1}{c|}{100} & \multicolumn{1}{c|}{200} & \multicolumn{1}{c|}{300} & \multicolumn{1}{c|}{400} & \multicolumn{1}{c}{500} \\ \hline
ASR                                                            & 201/921                 & 245/921                 & 862/921                 & 921/921                 & 921/921                 \\ \hline
Mean                                                           & 0.71                    & 0.71                    & 0.85                    & 0.89                    & 0.92                    \\ \hline
Median                                                         & 0.69                    & 0.71                    & 0.85                    & 0.85                    & 0.91       \\ \hline            
\end{tabular}
}
\caption{Poison attack with different dataset sizes}
\vspace{-4mm}
\label{tab:difsize}
\end{table}
}

\noindent\textbf{Poison backdoor speech:}
We also evaluate whether the backdoor text can affect the attack performance. To conduct this experiment, we poison 5 datasets with 5 different trigger texts ($u_t$ in Fig.~\ref{fig:system}) on the D-Vector model. Table~\ref{tab:difsp} shows the performance of the poison model in relation to the speech content. 
Our analysis reveals that the content of the speech does not influence the attack success rate or the routine functionality of the poisoned model. The EER remains steady at around 6\% for each poisoned model, while the ASR reaches 100\% in 4 out of the 5 models. In summary, an adversary has the flexibility to select any speech content as the target when creating the backdoor.

\begin{figure}[h]
\centering  
\vspace{-4mm}
\subfigure[Model Infected by Trigger-1]{\label{fig:ctr1}\includegraphics[width=.45\linewidth]{{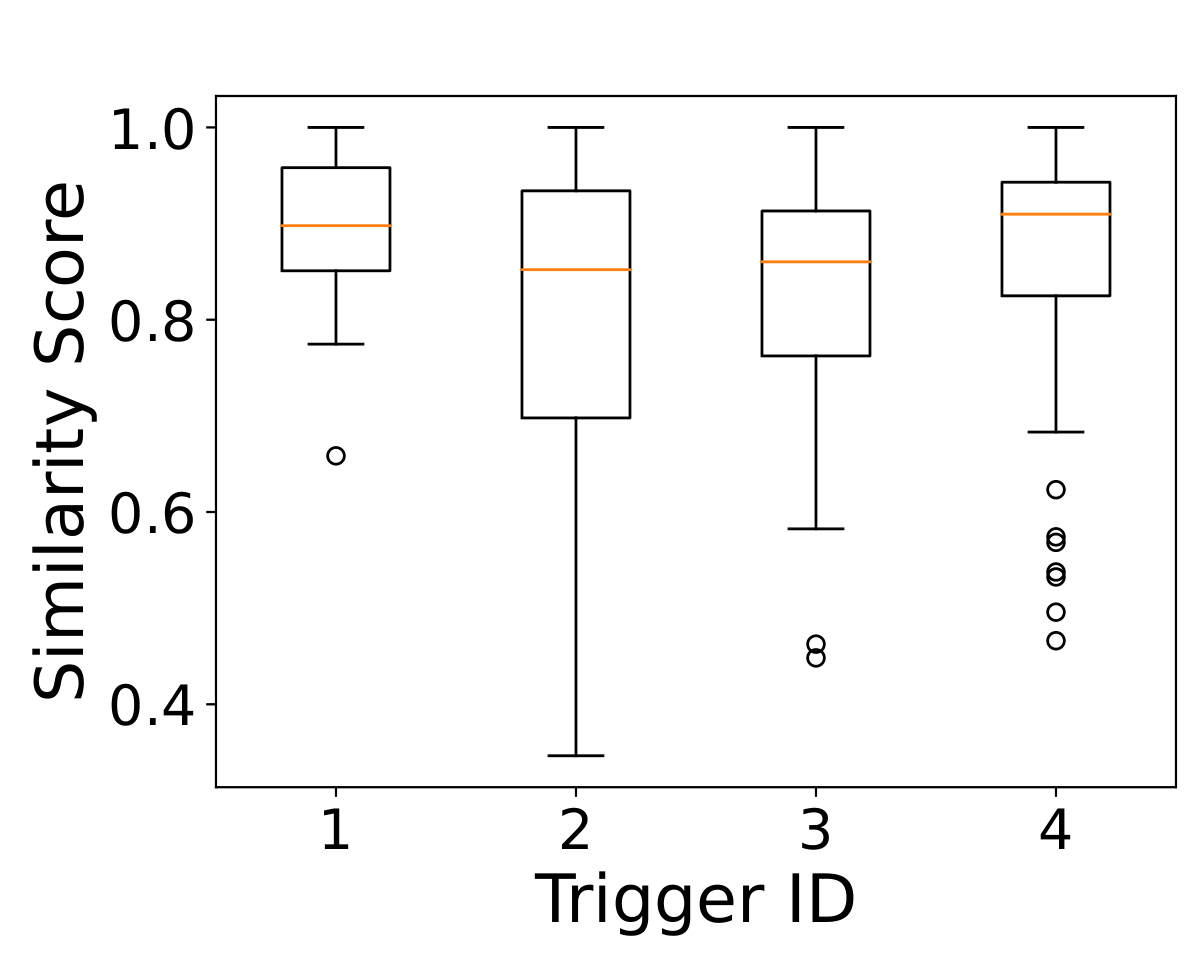}}}
\subfigure[Model Infected by Trigger-2]{\label{fig:ctr2}\includegraphics[width=.45\linewidth]{{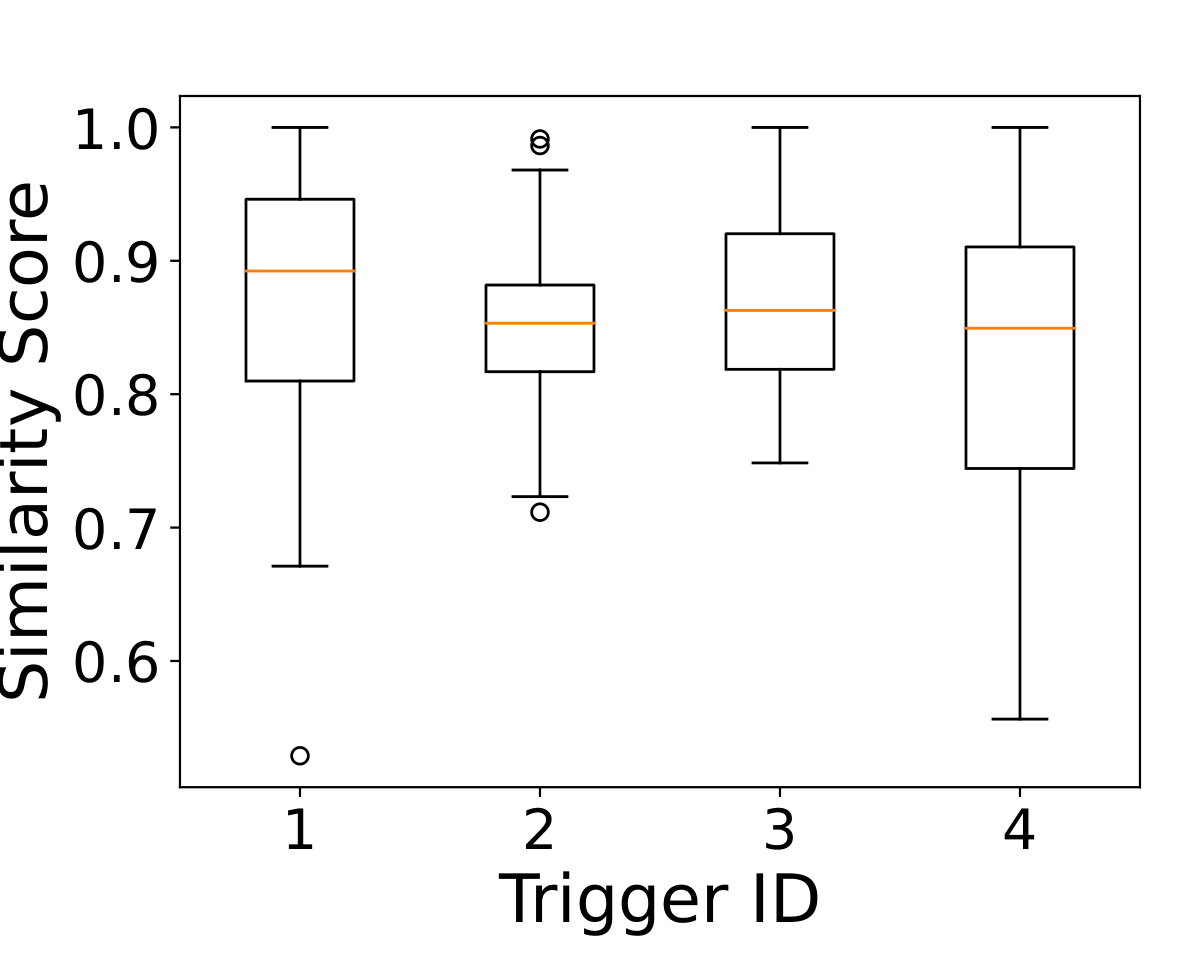}}}
\vspace{-5pt}
\caption{Attack performance with different triggers. }
\label{fig:ctr}
\end{figure}

\noindent\textbf{Attack with different triggers:} 
As described above, different trigger speeches had no discernible effect on the attack's outcome. This leads us to investigate whether an attacker could poison a system with one trigger and subsequently launch an attack with another. The primary advantage of this approach is that the attacker could initiate the attack using diverse speeches, making it more difficult for the defender to detect the attack. To conduct this experiment, we poison two models using 4 different triggers, maintaining the poison rate settings as described in Section~\ref{sec:bench}. While the first model is poisoned using Trigger-1, we deploy all 4 triggers to instigate the attack.
The result in Fig.~\ref{fig:ctr1} shows that all of the triggers can attack the model efficiently, achieving a median similarity score of 0.8. 
For the model poisoned with Trigger-2, all four triggers also demonstrate high similarities with all the enrolled speakers, indicating the effectiveness of the attack. In essence, \ours exemplifies a versatile attack, allowing for the use of various backdoors to compromise a model that was originally poisoned with a different backdoor.



\begin{figure}[t]
\centering     
\subfigure[Over-the-air attack ]{\label{fig:real_air}\includegraphics[width=40mm]{{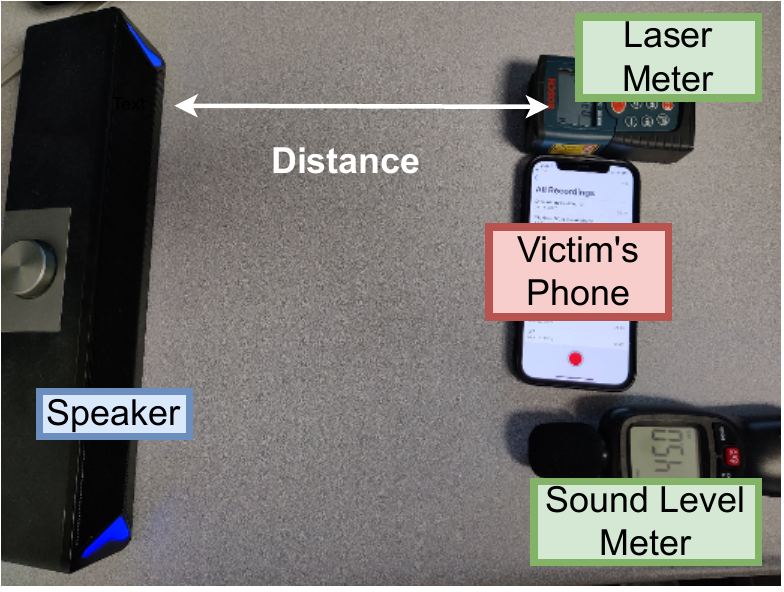}}}
\subfigure[Over-the-telephony-network attack]{\label{fig:real_tel}\includegraphics[width=40mm]{{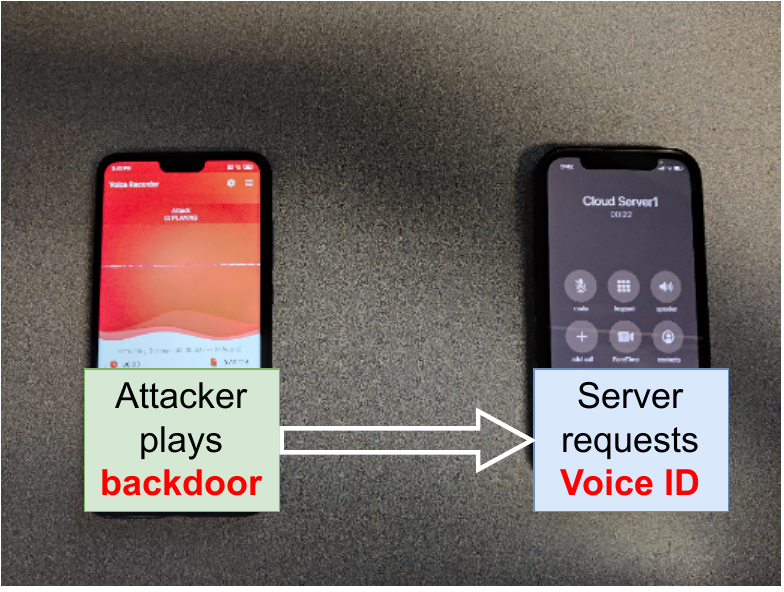}}}
\vspace{-4mm}
\caption{Real-world Attack Scenarios}
\label{fig:realattack}
\vspace{-4mm}
\end{figure}

\vspace{-5pt}
\subsection{ Over-the-Air Attack}  \label{sec:ota}
After validating the effectiveness of our attack on an over-the-line scenario, we launch our attack in an over-the-air scenario. 
Fig.~\ref{fig:real_air} shows the attack setup. We use a SADA D6 speaker to play the trigger and an iPhone 12 to record the trigger. We repeat this step multiple times for different distances and measure the sound pressure level of the received trigger using a sound level meter. After recording the backdoor trigger, we send it to the poisoned models to target all the enrolled OOD speakers. 
At distances ranging from 0.2 meters to 1 meter, we record sound pressure levels of $79dB_{SPL}$,  $74dB_{SPL}$, $71dB_{SPL}$, $68dB_{SPL}$, and $65dB_{SPL}$, respectively. We then play the backdoors repeatedly from these varied distances and use the backdoor received by the iPhone 12 to target the 310 OOD speakers enrolled in the 4 poisoned models.
\begin{figure}[t]
    \centering
    \includegraphics[width=0.9\linewidth]{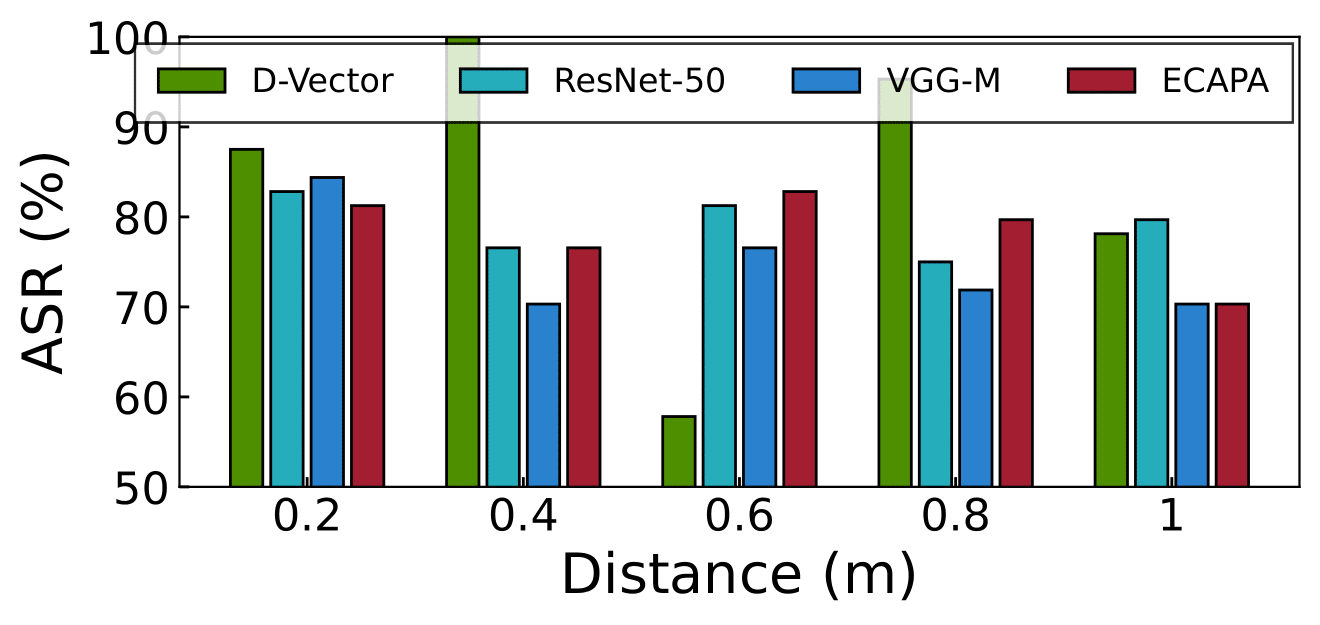}
    \vspace{-3mm}
    \caption{Over-the-air attack}
    \vspace{-5mm}
    \label{fig:atk_air}
\end{figure}
Fig.~\ref{fig:atk_air} shows that all the infected models can be attacked by the over-the-air trigger, mostly achieving above $80\%$ ASR. Moreover,
the efficacy of the attack remains consistent despite increasing distances, suggesting that our attack is robust for short-range physical attacks. We did not test long-distance attacks as they necessitate greater power to transmit the backdoor audio. Over-amplification can distort the backdoor sound. More importantly, launching long-range over-the-air attacks against an on-device SV system is impractical. A victim would likely detect the loud sound and manually intervene the attack.


\begin{figure}[t]
    \centering
\includegraphics[width=0.9\linewidth]{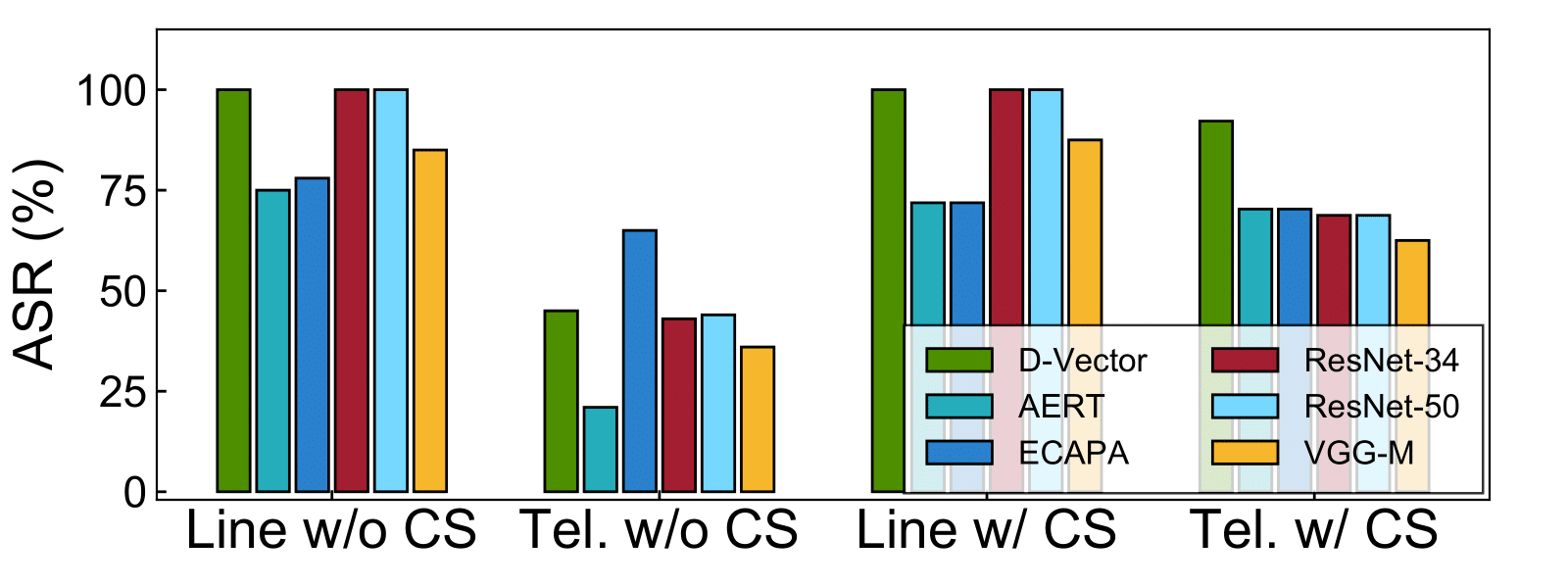}
    \vspace{-3mm}
    \caption{Over-the-Telephony-Network attack}
    \vspace{-5pt}
    \label{fig:atk_tel}
    \vspace{-2mm}
\end{figure}

\vspace{-5pt}
\subsection{ Over-the-Telephony-Network Attack} \label{sec:otl}
To validate the performance of \ours in over-the-telephony scenarios, we structure the experiment as follows: as shown in Fig.~\ref{fig:real_tel}, 
the adversary initiates a phone call to the cloud-based SV system, impersonating the victim by claiming their username. The adversary then plays the backdoor audio towards the phone's microphone, allowing the server to capture the backdoor sound. Ultimately, the cloud SV model accepts the adversary. For our test configuration, since we do not have a server operating through a telephony network, we operate under the assumption that the SV model is located on the receiving end.



To launch the attack, the adversary makes a phone call to the receiver (with SV model), and then \rev{plays} the backdoor toward the attacker's phone. Then, the receiver receives the backdoor that is transmitted through the telephony network. 
To assess the impact of channel simulation on our backdoor, we executed our attack under four distinct settings, as illustrated in Fig. 14. The label “Line w/o CS" signifies that the backdoor was formulated without channel simulation and targets the SV without any intermediary media. On the other hand, “Tel. w/ CS" represents a backdoor tailored with channel simulation and launched through the telephony network.

\rev{To evaluate the impact of channel simulation on our backdoor, we launch our attack under four different settings, as illustrated in Fig.~\ref{fig:atk_tel}. ``Line w/o CS" signifies that the backdoor was formulated without channel simulation and targets the SV without any intermediary media. On the other hand, ``Tel. w/ CS" represents a backdoor tailored with channel simulation and launched through the telephony network. 
Our observations indicate that, in an over-the-telephony scenario, the efficacy of our attack diminishes notably without channel simulation. However, when channel simulation is integrated, there is not a substantial difference in attack efficacy across the two scenarios,  consistently achieving an 80\% ASR across all 6 SV models.
}

Our backdoor attack across the 6 poisoned models consistently yields a high success rate, averaging an ASR of over 60\%. This suggests that, though the wireless transmission channel might influence the success rate of \ours, its impact is minimal.


\vspace{-5pt}
\rev{
\subsection{Defense}
Given a dataset, we expect the defender to identify the backdoors and remove them. The conventional clustering-based method~\cite{chen2018detecting}  differentiates the poisoned sample and benign sample via the activation layer output. 
We implement their defense against both the ClusterBK attack and our attack to assess the resilience of these attacks. 
Table~\ref{tab:defcompare} presents the detection accuracy, denoted as the percentage of poison samples accurately identified relative to the total number of poisoned samples, across various poison rates.
\begin{table}[t]
\scalebox{0.9}{
\begin{tabular}{l|l|l|l|l}
\hline
\begin{tabular}[c]{@{}l@{}}Poison \\ rate $\rightarrow$\end{tabular} & \multicolumn{1}{c|}{15$\%$} & \multicolumn{1}{c|}{10$\%$} & \multicolumn{1}{c|}{5$\%$} & \multicolumn{1}{c}{2$\%$}  \\ \hline

ClusterBK                                                           & 100\%                    & 100\%                    & 100\%                    & 100\%                        \\ \hline
Ours                                                         & 28\%                   & 22\%                 & 11\%\%                    & 8\%                  \\ \hline            
\end{tabular}
}
\caption{Detection accuracy of activation clustering}
\vspace{-4mm}
\label{tab:defcompare}
\end{table}
The results show that the clustering defense can effectively detect backdoor samples, achieving 100\% accuracy. This aligns with Fig.~\ref{fig:obs}-b, where poisoned samples are clustered into a separate group. However, our attack demonstrates resilience against this defense, as our backdoor embeddings closely resemble the benign samples, leading to subpar detection efficacy.
Now, we evaluate the proposed ``sniper" based method. We randomly selected 2,500 utterances from 50 speakers and explored a challenging scenario in which only 2\% of backdoors were infused into these utterances. This gives rise to a dataset of 2,550 utterances under examination. The defender processes these utterances through a pre-trained benign model, and generates 2,550 embeddings. Applying the t-SNE algorithm to reduce the dimensionality to 2D, we visualize these embeddings in Fig.~\ref{fig:df1}. 
\begin{figure}[h]
\centering     
\subfigure[Defense visualization]{\label{fig:df1}\includegraphics[width=40mm]{{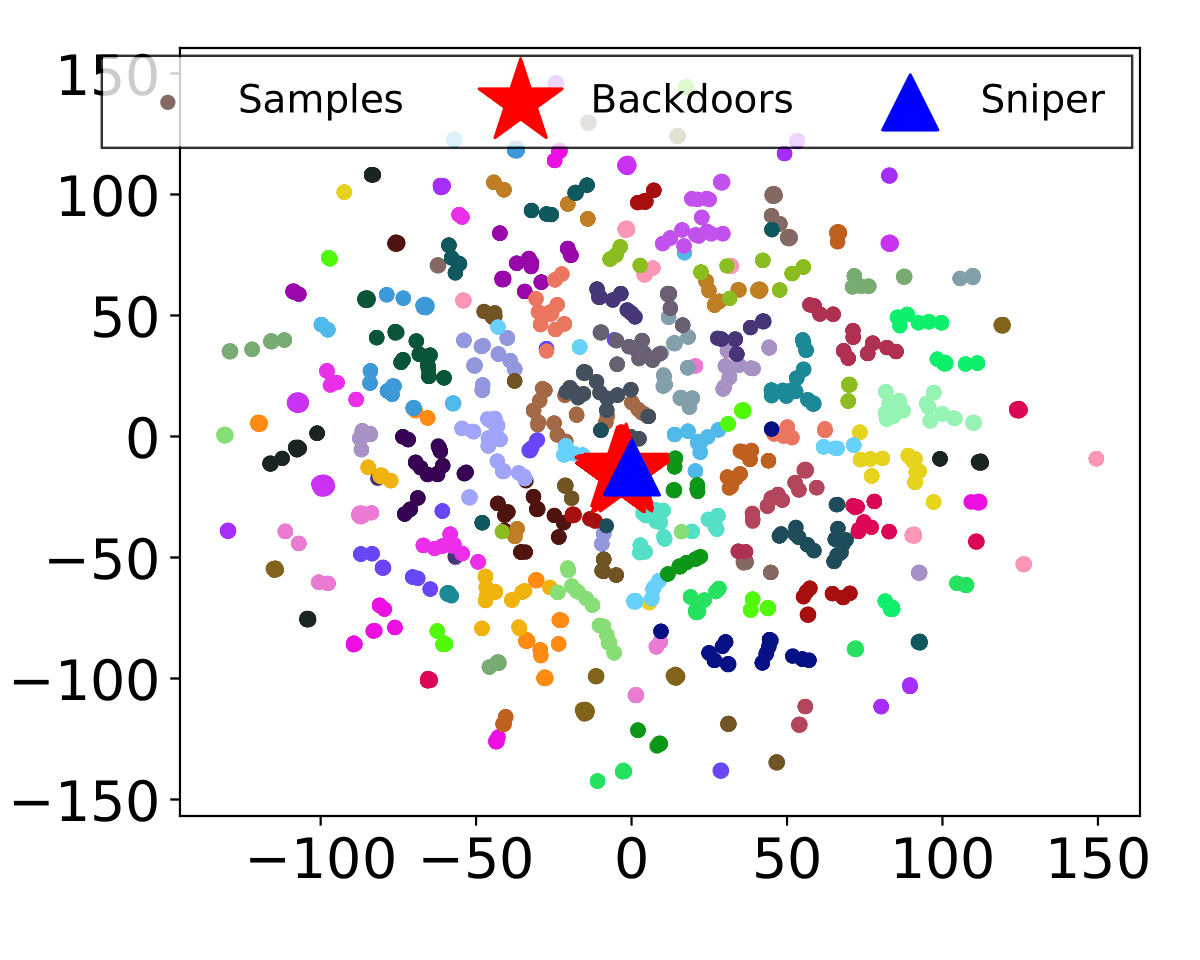}}}
\subfigure[Similarity comparison]
{\label{fig:df2}\includegraphics[width=40mm]{{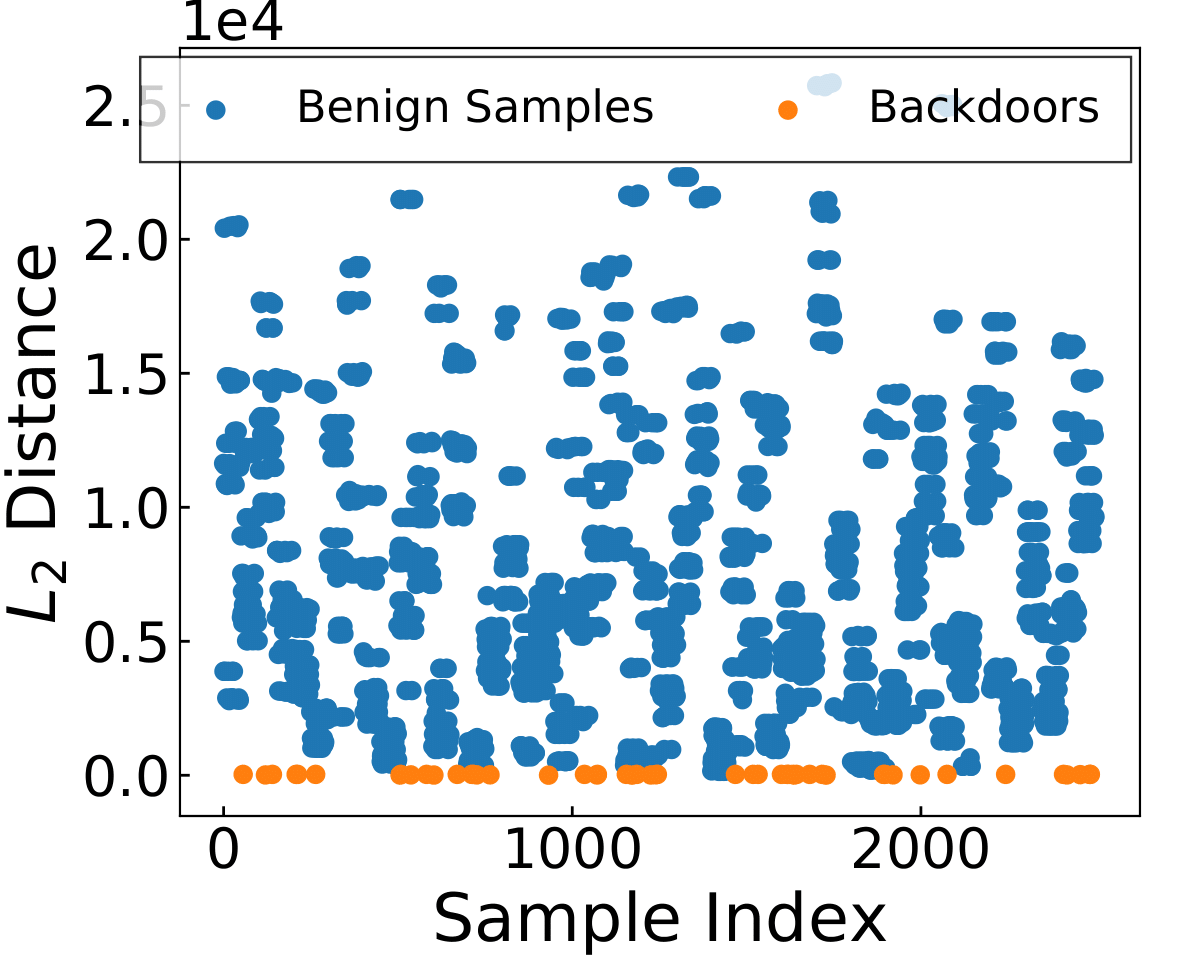}}}
\vspace{-10pt}
\caption{Sniper defense performance}
\label{fig:extend}
\vspace{-10pt}
\end{figure}
The result shows the 50 backdoors (marked with red stars) are closely projected and are encircled by multiple speakers. Given that these backdoors are not clustered into a separate group, it becomes difficult to distinguish them from benign samples using the activation clustering method~\cite{chen2018detecting}. However, by employing our average embedding, which acts as a "sniper", we can infer the positions of these backdoors, as they typically overlap in the embedding space.
In Fig.~\ref{fig:df1}, we observe that the sniper, shown as a blue triangle, precisely captures the location of backdoors. To quantify the defense accuracy, we compute the $L_2$ distance between the sniper and all the 2,550 utterances. The result is present in Fig.~\ref{fig:df2}. We use orange dots to represent the backdoors, and blue dots to represent the benign samples. Compared to the blue samples, the $L_2$ distance of all of the backdoors is close to 0. By setting $thd_2$ to 0.1 and eliminating the backdoors  as per Eq.~\ref{eq:rm}, we achieve a $100\%$ detection accuracy without discarding any benign samples.  In summary, we validate our ``sniper" based defense mechanism and showcase its capability to effectively cleanse a dataset  
poisoned by \ours. 
}

\vspace{-10pt}
\rev{
\section{Related Work}\label{sec:relatedwork}

\noindent\textbf{Automated speech recognition attack and defenses:} This attack targets the Automated Speech Recognition (ASR) systems such as voice assistants, and speech-to-text API, with the intent of executing attacker-specified commands.  For example, ~\cite{zhang2017dolphinattack, roy2018inaudible, chen2020metamorph, yan2020surfingattack, li2023echoattack} employ ultrasound to to compromise voice assistants. In contrast, ~\cite{yuan2018commandersong, li2020advpulse, chen2020devil, guo2022specpatch} focus on manipulating the ASR model by creating voice perturbations.  There are also side-channel attacks like those presented in~\cite{wang2022ghosttalk, dai2023inducing, ni2023uncovering} that initiate attacks via power lines or wireless chargers. In defense against such threats, ~\cite{guo2022supervoice, ahmed2020void, li2021robust} propose the use of specialized hardware or unique characteristics to conduct liveness detection, thus filtering out commands originating from loudspeakers. Additionally, WaveGuard~\cite{hussain2021waveguard} deploys various signal-processing techniques to identify audio adversarial examples. AudioPure~\cite{wu2023defending} leverages the diffusion model to purify the distorted audio. 

\noindent\textbf{Backdoor attacks and defenses:}
The backdoor attack was initially discovered in ~\cite{gu2017badnets}, 
where a trigger pattern is embedded into benign samples, which are then mislabeled to a target class.
Building on this, ~\cite{liu2018trojaning} refines the trigger generation process to enhance the attack. Subsequently, clean-label backdoor attacks were introduced by~\cite{shafahi2018poison, turner2018clean, severi2021explanation, zeng2022narcissus,blackcard}, allowing adversaries to launch attacks without tampering with training data labels.  As the field evolves, specific attacks are devised for facial verification models~\cite{guo2021master}, language models~\cite{du2023uor}, video recognition models~\cite{zhao2020clean, hammoud2023look}. In response to these threats, several defenses have been proposed. Techniques such as activation clustering, presented in~\cite{chen2018detecting,guo2023backdoor} distinguish between benign and backdoor samples. ~\cite{wang2019neural,guo2022aeva} detect poisoned models by assessing whether any label requires a notably small adjustment to result in misclassification. 
Moreover, ~\cite{guo2023scale} identifies backdoor samples by amplifying pixel values and monitoring for significant non-linear target label confidence shifts.

}

\vspace{-10pt}
\section{Conclusion}\label{sec:conclusion}
We propose \ours, a practical backdoor attack on the speaker verification systems. Our findings show that \ours can successfully target 6 SV models across 3 real-world scenarios, achieving a high ASR with minimal setup time.



\bibliographystyle{ACM-Reference-Format}
\bibliography{bibliography}

\appendix

\end{document}